\documentclass{jfm}
\usepackage{graphicx}
\usepackage{epstopdf, epsfig}
\usepackage{amsfonts,amssymb,amsmath,mathbbol}   

\usepackage{color}
\usepackage{multirow}
\usepackage{dcolumn}
\usepackage{multicol}
\usepackage{graphicx}
\usepackage{caption}
\usepackage{amssymb}
\usepackage[normalem]{ulem}
\usepackage{amsmath}
\usepackage{epstopdf}
\usepackage{mathtools}
\usepackage{float}
\usepackage{bm}
\usepackage{ulem}
\usepackage{booktabs}
\usepackage{subcaption}
\usepackage{graphicx}
\usepackage{soul}
\usepackage{color}
\usepackage{xcolor} 
\usepackage{hyperref}
\usepackage{placeins}

\def\We{{\it We}}

\usepackage{booktabs, makecell, multirow}

\renewcommand{\arraystretch}{1.7}

\usepackage{siunitx}
\shorttitle{Submerged Immiscible Oil Jet}
\shortauthor{L. M. Sharma, A. Aiyer, H. N. Dixit and L. D. Chandrala}

\title{Submerged Immiscible Oil Jet: Breakup Dynamics and Droplet Size Modelling}

\author{
Lokendra Mohan Sharma\aff{1},
Aditya Aiyer\aff{2},
Harish N. Dixit\aff{1,3}
\and
Lakshmana Dora Chandrala\aff{1}\corresp{\email{lchandrala@mae.iith.ac.in}}
}

\affiliation{
\aff{1} Department of Mechanical and Aerospace Engineering,  
Indian Institute of Technology Hyderabad,  
Kandi-502~284, Sangareddy, Telangana, India

\aff{2} Department of Mechanical Engineering and Mechanics, 
Lehigh University, 
Bethlehem, PA, United States

\aff{3} Centre for Interdisciplinary Programs,  
Indian Institute of Technology Hyderabad,  
Kandi-502~284, Sangareddy, Telangana, India
}

\begin{document}
\maketitle

\begin{abstract}
Immiscible liquid-liquid jets occur in numerous industrial and environmental settings, including subsea oil spill response and liquid-liquid extraction, yet the coupled evolution of turbulence, interfacial instability, and droplet size distribution across transitional to turbulent regimes remains poorly understood. We present a combined experimental and theoretical investigation of a silicone oil jet injected into refractive-index-matched sugar water, employing simultaneous Particle Image Velocimetry (PIV) and Planar Laser-Induced Fluorescence (PLIF). Our results indicate that, beyond the breakup region, mean velocity and Reynolds stress profiles exhibit self-similarity. Proper Orthogonal Decomposition (POD) of the volume fraction reveals a systematic shift in the dominant instability mechanism with increasing Reynolds number: axisymmetric Rayleigh–Plateau modes govern breakup at low Reynolds numbers, sinuous and varicose modes coexist in the transitional regime, and Kelvin–Helmholtz-driven shear-layer breakup dominates at higher Reynolds numbers. Droplet sizes are extracted from PLIF images using a convolutional neural network-based segmentation approach. In the transitional regime, a bimodal droplet size distribution is observed, reflecting the coexistence of shear-induced breakup and Rayleigh–Plateau instability. As the Reynolds number increases and turbulent dissipation intensifies, the distribution evolves toward a single mode with smaller characteristic droplet sizes. To predict the droplet size distribution, existing population balance models are extended to include droplet coalescence. Incorporating coalescence substantially improves agreement with experimental measurements, particularly for larger droplets. These findings demonstrate that turbulent dissipation, interfacial instabilities, and droplet coalescence collectively govern the evolution of droplet size distribution in immiscible liquid-liquid jets.
\end{abstract}


\section{Introduction} \label{sec:intro}
Multiphase jet flows involving immiscible liquid-liquid systems are encountered across a broad range of industrial, environmental, and geophysical settings, including emulsification, liquid-liquid extraction, chemical reactor design, and subsea oil spill response \citep{zheng2003model, johansen2013droplet, socolofsky2011formation, yang2016large}. The interplay of inertia, interfacial tension, viscous stresses, and buoyancy dictates how an injected liquid jet deforms, fragments, and forms droplets. A notable application is the subsurface release of crude oil during blowout events \citep{socolofsky2011formation,fingas2016oil,ramseur2010deepwater}, where the size distribution of the fragmented oil droplets determines their subsequent fate. The larger droplets rise buoyantly to the sea surface, while smaller droplets remain suspended in the water column, where they are subject to turbulent transport and microbial degradation by marine bacteria \citep{head2006marine}. Accurate prediction of the droplet size distributions and the underlying breakup mechanisms is therefore of direct practical importance for assessing the environmental impact of oil spills and designing effective response strategies. Despite this significance, understanding the mechanisms by which inertia, interfacial tension, and turbulence interact to produce a polydisperse droplet population remains an open problem, particularly in the turbulent regimes where predictive modelling is most challenging.
 
The breakup regime of a liquid jet is characterised by the Reynolds number $\Rey = \rho_d V_{mean} D / \mu_d$, the Weber number $\We = \rho_c V_{mean}^2 D / \sigma$, and the Ohnesorge number $Oh  = \mu_d / \sqrt{\rho_d \sigma D}$, where $\rho_d$, $\rho_c$, $\mu_d$, $V_{mean}$, and $D$ denote the jet density, continuous phase density,  dynamic viscosity of jet, mean injection velocity, and nozzle diameter, respectively, and $\sigma$ is the interfacial tension. While these parameters suffice to characterise single-phase jet breakup \citep{machicoane2019synchrotron}, immiscible liquid-liquid systems introduce two additional dimensionless groups, the density ratio $\rho_d/\rho_c$ and the viscosity ratio $\mu_d/\mu_c$, where subscript $c$ denotes the continuous phase.
At low Reynolds numbers, surface tension drives the classical Rayleigh-Plateau (RP)  instability, yielding nearly monodisperse droplets whose size is governed by the most amplified capillary wavelength \citep{rayleigh1878instability, saito2017flow, sharma2025scaling}. At high Reynolds numbers, Kelvin-Helmholtz (KH) instabilities dominate the shear layer, driving a turbulent  atomisation cascade that produces a highly polydisperse droplet population. While practically important, there is a lack of systematic data on how the droplet size distribution evolves across these regimes in immiscible liquid-liquid systems.

A key challenge in studying immiscible jet breakup experimentally is the optical aberrations arising from the refractive index mismatch between the two liquid phases. In practical systems, such as crude oil in seawater, the opacity of the dispersed phase severely limits optical access, while even in transparent oil-water systems,  refractive index differences at the jet interface introduce optical aberrations that distort planar imaging measurements such as Particle Image Velocimetry (PIV) and Planar Laser-Induced Fluorescence (PLIF). These constraints have motivated the use of refractive index-matched fluid pairs that preserve the relevant physical properties while enabling simultaneous, quantitative measurements of the velocity field and dispersed-phase concentration. Exploiting this approach, \citet{xue2021flow} applied simultaneous PIV and PLIF to characterise the mean and fluctuating fields of an immiscible buoyant oil jet, demonstrating that turbulence production is strongly phase-dependent and that droplet breakup continues well beyond the near-field region.  Along similar lines, \citet{kewalramani2022experimental}  employed a refractive-index-matched liquid--liquid system together with PLIF imaging to investigate the Lagrangian evolution of droplets generated from turbulent immiscible jets. Their measurements quantified droplet size, velocity, breakup, and collision events, demonstrating the simultaneous occurrence of breakup and coalescence during jet dispersion. However, their study primarily focused on dispersed droplet dynamics after breakup and did not examine the evolution of the turbulent flow field or its coupling with droplet size distribution. \citet{johansen2013droplet} conducted oil jet breakup experiments in a tower tank facility and measured the far-field droplet size distribution using Laser In-Situ Scattering and Transmissometry. Based on these measurements, they developed an empirical model that relates the volumetric mean droplet size to a modified Weber number incorporating viscous resistance to breakup. The resulting droplet size distribution was found to be well described by a Rosin–Rammler function. \citet{ibarra2020near} showed that interfacial length scales in immiscible turbulent jets decrease with Reynolds number, and that droplet size is governed primarily by the competition between turbulence intensity and interfacial tension. \citet{campbell2020jet} identified distinct jetting regimes in liquid-liquid Taylor vortex flow and demonstrated that mean droplet size correlates primarily with the jet  Reynolds number.  Despite these contributions, quantitative data on the turbulence structures, instability transition, and evolving droplet size distribution across a wide range of Reynolds numbers remain limited for immiscible liquid-liquid jets. Recently, \citet{rimbert2023mass} combined experiments and theoretical modelling to investigate liquid--liquid jet fragmentation. They showed that the fragment mass distribution can be represented by a log-stable law arising from a two-step fragmentation cascade involving viscous shear instability followed by Rayleigh--Taylor instability. Their work provides a physics-based interpretation of fragment size distributions, highlighting the role of instability mechanisms in determining the final droplet population.
 
On the modelling side, accurate prediction of the droplet size distribution in turbulent multiphase flows is essential for estimating mass transfer rates, biodegradation timescales, and transport pathways in both natural and engineered systems. Population balance models (PBMs) have emerged as the principal framework for this purpose, describing the temporal evolution of the droplet number density through source terms that account for breakup and coalescence. Early phenomenological models by \citet{coulaloglou1977description} and \citet{tsouris1994breakage} treated droplet-eddy interactions by analogy with kinetic theory, introducing breakup frequency and efficiency functions based on the energy balance between turbulent fluctuations and interfacial resistance. These models were extended by \citet{zhao2014vdrop} to incorporate viscous resistance to breakup in the theoretical framework, which introduced a time-dependent viscous resistance term accounting for the elongation dynamics of high-viscosity droplets. \citet{aiyer2019population} developed a comprehensive large eddy simulation (LES) model coupled with a population balance equation, extending existing breakup frequency models to account for both inertial and viscous range turbulence through a second-order structure function formulation. Building on this, \citet{aiyer2020coupled} introduced a one-dimensional (1D) ordinary differential equation (ODE) model that predicts the evolution of the droplet size distribution along the jet centreline by accounting for axial advection, radial turbulent transport, and droplet breakup, providing a computationally efficient far-field prediction. These modelling developments, however, have focused primarily on breakup-dominated regimes, and the additional role of droplet coalescence, which becomes significant when the dispersed-phase concentration is elevated and turbulent dissipation is reduced, has not been consistently incorporated or experimentally validated in the context of liquid-liquid jets.  Interface-resolving methods such as direct numerical simulation (DNS) with volume-of-fluid or level-set techniques can capture individual droplet deformation and breakup, but their computational cost escalates prohibitively when droplet sizes span orders of magnitude smaller than the nozzle diameter \citep{desjardins2008accurate, ling2019two}, reinforcing the importance of reduced-order population balance models informed and validated by reliable experimental data.
 
In addition to this, the detailed spatio-temporal dynamics of interfacial deformation and the emergence of coherent flow structures during the breakup of immiscible jets are not fully understood. In particular, the coupling between coherent vortical structures and droplet formation processes has received limited quantitative attention. Data-driven decomposition techniques, such as Proper Orthogonal Decomposition (POD) \citep{sirovich1987turbulence}, offer a powerful framework to extract dominant flow modes and their temporal evolution from experimental data, thereby linking coherent motions to breakup dynamics.
 
In the present study, we address these gaps through a combined experimental and theoretical investigation of immiscible liquid-liquid jet breakup over the Reynolds number range $Re = 600$-$1700$, spanning the transitional and  turbulent regimes. A refractive index-matched silicone oil-sugar water fluid pair is employed to enable simultaneous, phase-resolved PIV and PLIF measurements of the velocity field and dispersed-phase volume fraction. The jet dynamics are characterised through mean flow evolution, modal decomposition of the PLIF snapshots using proper orthogonal decomposition (POD), and turbulence statistics obtained from PIV. Droplet sizes are extracted from the PLIF images using a semantic segmentation approach based on the U-Net convolutional neural network (CNN) architecture \citep{ronneberger2015u}, enabling robust detection of individual droplets across a wide range of sizes in the presence of out-of-focus background structures. The measured droplet size distributions are compared with predictions from the one-dimensional population balance model of \citet{aiyer2020coupled}, with coefficients determined from the experimental data. The model is further extended to incorporate droplet coalescence, and the relative contributions of breakup and coalescence to the evolving droplet size distribution are quantified.
 
The remainder of the paper is organised as follows. Section~\ref{sec:expt} describes the experimental setup, including the fluid pair selection, optical measurement system, and image processing methodology.  The results and discussion are presented across two sections. Section~\ref{sec:results} presents the flow characterisation and droplet size measurements, covering jet dynamics and instability mechanisms characterised through PLIF imaging and POD modal analysis, turbulent flow statistics from PIV, and droplet size distributions extracted via CNN-based segmentation. Section~\ref{sec:pbm} presents the population balance model formulation, its validation against experimental measurements, and a parametric study of the effect of continuous phase viscosity. Concluding remarks are given in Section~\ref{sec:conclusion}.

\section{Experimental setup}
\label{sec:expt}
\begin{figure}
\centering
\includegraphics[width=0.85\textwidth]{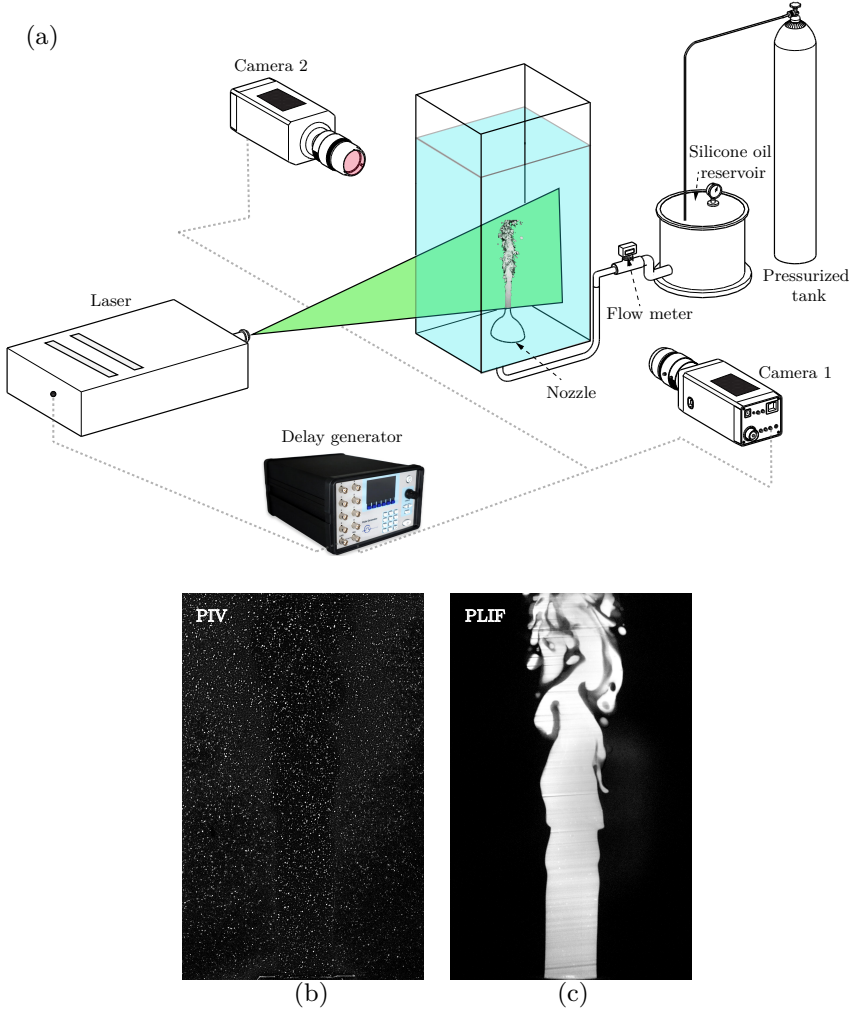}
\caption{(a) Schematic representation of the experimental setup illustrating the simultaneous measurement of the interface position using PLIF and the velocity field using PIV. Sample images from the experiments are shown: (b) PIV image acquired using camera-1 with a band-pass filter and pre-processed by background subtraction and histogram  equalisation, and (c) corresponding PLIF image acquired by camera-2 equipped with a long-pass filter. }
\label{fig:exp_setup}
\end{figure}

The schematic of the experimental setup is shown in figure~\ref{fig:exp_setup} (a). The tank is constructed from acrylic to facilitate optical access. To minimise recirculation, the tank cross-section was designed to be 30 times the nozzle diameter and the height 100 times the nozzle diameter. Silicone oil (polydimethylsiloxane) was injected into the tank from the bottom through a nozzle of inner diameter $D = 10$~mm connected to a pressurised reservoir, with the flow rate continuously monitored using an ultrasonic flow meter. The nozzle assembly incorporates a splitter plate, honeycomb, mesh grid, and a seventh-order contraction to produce a uniform top-hat velocity profile at the nozzle exit. 
 
Probing planar velocity fields and volume fractions in immiscible liquid--liquid systems is challenging due to optical aberrations caused by refractive index mismatch between the phases. The silicone oil used in this study has a refractive index of 1.4004, compared with 1.33 for distilled water. To eliminate these aberrations, a refractive index-matched fluid pair was employed following \cite{xue2021flow} and \cite{sharma2025scaling}. Sugar ($\text{C}_{12}\text{H}_{22}\text{O}_{11}$) was incrementally dissolved in distilled water, and the refractive index was monitored at each step using a refractometer (Make: Atago, Model: PAL-RI). A concentration of 39\% weight/volume was found to raise the refractive index of the solution to 1.4002, closely matching that of the silicone oil. The interfacial tension was measured using a Kr\"{u}ss contact angle system in pendant drop configuration, in which a silicone oil droplet was suspended upward into the sugar water solution through a needle inserted from the bottom of a small cuvette. The viscosities of the fluids were measured using an Anton Paar rheometer  (Make: Anton Paar, Model: IC201-3). The physical properties of both fluids are summarised in Table~\ref{table:properties}.

\begin{table}
    \centering
    \caption{ Physical properties of  the selected liquid pair}
    \label{table:properties}
\begin{tabular}{@{} ccccc @{}}
    \toprule
\thead{Fluid}
        & \thead{ $\rho~(\mathrm{kg\,m^{-3}})$}
                    & \thead{ $\mu~(\mathrm{kg\,m^{-1}\,s^{-1}})$} 
                          &\thead{$RI$}
                                 &\thead{ $\sigma~(\mathrm{N\,m^{-1}})$}\\
    \midrule
Silicone oil   & 960              & $(9.6 \pm 0.2) \times 10^{-3}$       & 1.4004          &      \multirow{2}{*}{0.036}         \\
Sugar water (39 \% w/v)   & 1150             & $(5.65 \pm 0.2) \times 10^{-3}$       & 1.4002          &               \\
 
    \bottomrule
\end{tabular}
    \end{table} 
 
The optical diagnostics comprises of a high-speed Nd:YLF laser (Model: DM30-527; Photonics Industries Inc., USA), two high-speed cameras (Model: Phantom VEO 640L; Vision Research, USA), and a digital delay generator. The flow is illuminated by a planar laser sheet of approximately 0.5~mm thickness, generated by a sheet optics assembly consisting of cylindrical and spherical lenses. The two high-speed cameras are arranged on either side of the tank, as shown in figure~\ref{fig:exp_setup}(a), to enable simultaneous PIV and PLIF measurements. Both cameras use lenses of identical focal length (Tokina Macro lens, f= 100 mm) at equal working distance from the measurement plane, and are calibrated using a dual-plane target (black background, white dot grid) of known thickness. Since the laser sheet is aligned to one face of the target, only one camera is initially in focus; the target is then translated along the optical axis by a micrometer traverse, by a distance equal to its thickness, so that both cameras become focused on the same physical plane. The resulting PIV and PLIF coordinate systems are each shifted to the nozzle centreline to have a common reference frame. Nanosecond synchronisation of the laser pulses and camera exposures is achieved using the digital delay generator. For PIV, the sugar water (continuous phase) is seeded with hollow glass spheres of 10~$\mu$m diameter, and the dispersed silicone oil phase  is seeded with polyamide particles of 5~$\mu$m diameter. To visualise the dispersed phase boundaries, the oil-soluble fluorescent dye Nile Red is dissolved in the silicone oil. The dye has an excitation range of 515--559~nm and an emission range of 580--640~nm. The Mie scattering and fluorescence signals are separated by fitting a narrow bandpass filter (532~nm $\pm$ 5~nm) to the PIV camera lens and a longpass filter (560~nm) to the PLIF camera lens. The time between the laser pulses is set to 500~$\mu$s, 
ensuring a particle displacement of approximately 25\% of the interrogation window length. Both cameras operate at 1000~fps with a resolution of 2560 $\times$ 1600 pixels.  The macro lenses mounted on each camera yield a physical field of view of 140~mm $\times$ 87~mm. Both cameras share the same raw image spatial resolution of approximately 18~px/mm, since identical lenses and the same physical field of view are used. However, the vector resolution of the PIV is coarser: the final interrogation window size is $16 \times 8$~pixels (axial $\times$ transverse) with 50\% overlap, yielding a PIV vector spacing of $8 \times 4$~pixels, corresponding to approximately $0.44~\text{mm} \times 0.22$~mm. The PLIF images retain the full pixel-level spatial resolution of $\approx 0.056$~mm/px, which is used directly for the droplet size analysis.

\begin{figure}
\centering
\includegraphics[width=0.8\textwidth]{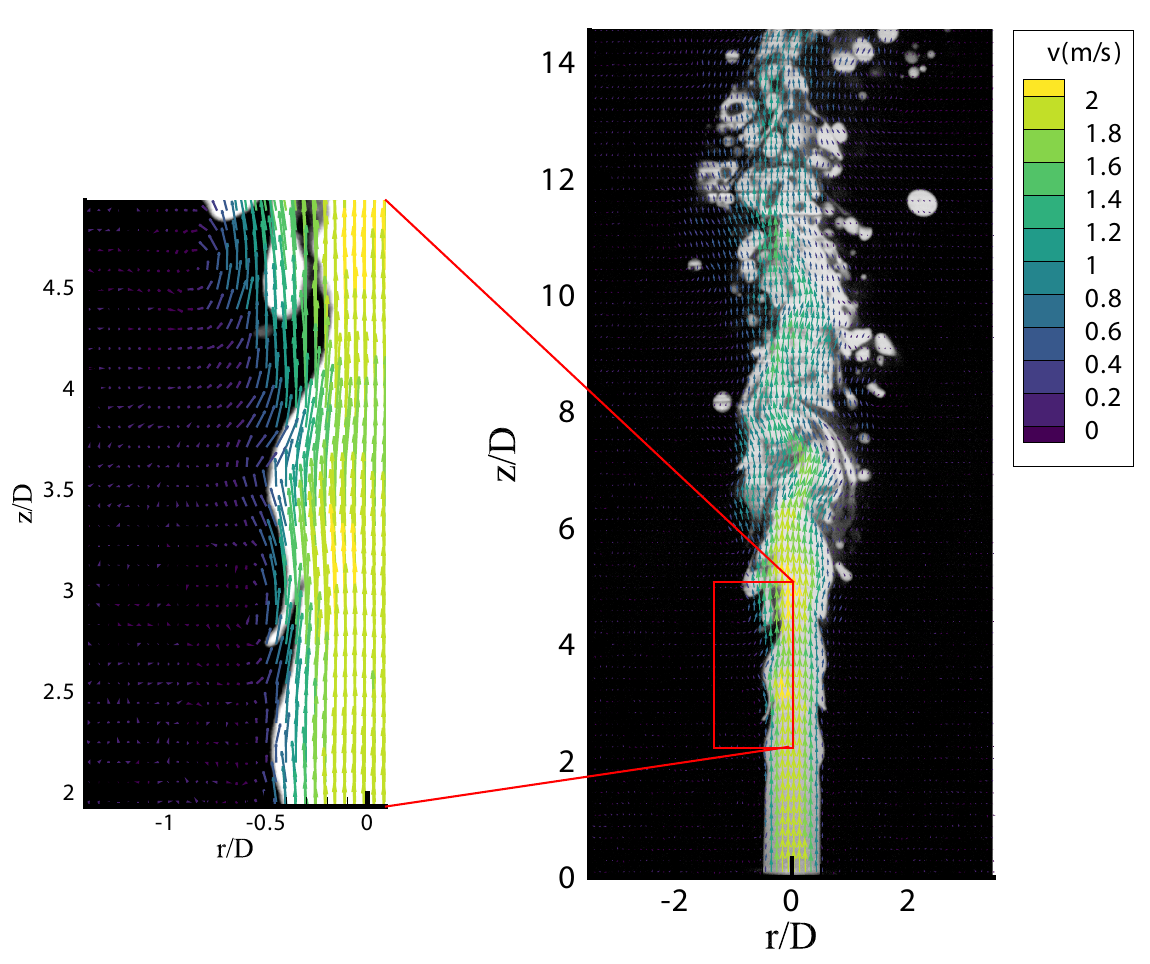}
\caption{Instantaneous PLIF image of the silicone oil jet in sugar water  overlapped with velocity vectors obtained from PIV for $Re  =1400$. The contour indicates the magnitude of axial velocity. The inset shows an enlarged view of the region near the Kelvin-Helmholtz roll-up. }
\label{fig:instanteneous_vel}
\end{figure}

Before cross-correlation analysis, the PIV images are pre-processed to enhance the signal-to-noise ratio. The residual fluorescence from Nile Red emission is removed by subtracting a background image obtained by median filtering the original image, followed by histogram equalisation to enhance particle contrast~\citep{sharma2025scaling}. Representative pre-processed PIV and PLIF images are shown in figures~\ref{fig:exp_setup}(b) and (c), respectively. 
A total of 4000 velocity maps are ensemble-averaged to obtain the mean velocity field. The detailed procedure for extracting droplet size distributions from these images is described in Section~\ref{sec:dsd}. The uncertainty in the PIV velocity measurements was estimated following \citet{forliti2000bias} and found to be within 0.005~m/s. Each experiment was repeated three times, and the standard deviation across realisations is reported as error bars. A representative instantaneous velocity field superimposed on the corresponding PLIF image for $\Rey = 1400$ is shown in figure~\ref{fig:instanteneous_vel}. The enlarged inset highlights the velocity vectors in the jet shear layer, revealing the local flow structure and developing interfacial instabilities.

\section{Flow Characterisation and Droplet Size Measurements}
\label{sec:results}
This section is organised as follows. Sections~\ref{sec:jet_dynamics} and~\ref{sec:pod_plif} characterise the jet dynamics and dominant instability mechanisms through PLIF imaging and POD modal analysis. Section~\ref{sec:turbulent_characteristics} analyses the turbulent flow statistics derived from PIV measurements. Section~\ref{sec:phase_conditioned} presents the phase-conditioned turbulent statistics. Section~\ref{sec:dsd} presents the experimentally measured droplet size distributions.

\subsection{Characterisation of Jet Dynamics}
\label{sec:jet_dynamics}
\begin{table}
\centering
\caption{Non-dimensional parameters corresponding to different experimental cases.}
\label{table:ndn}
\begin{tabular}{cccc}
\toprule
Cases & $\Rey$ & $We$ & $Oh$ \\
\midrule
Case~0 & 125  & 6.64  & \multirow{5}{*}{0.0163} \\
Case~1 & 600  & 153  & \\
Case~2 & 1000 & 425  &  \\
Case~3 & 1400 & 834  &  \\
Case~4 & 1700 & 1229 &  \\
\bottomrule
\end{tabular}
\end{table}

\begin{figure}
\centering
\includegraphics[width=0.8\textwidth]{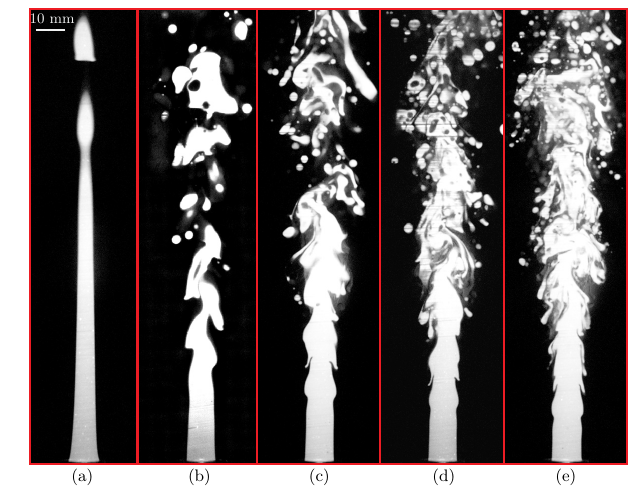}
\caption{Instantaneous PLIF images of the silicone oil jet in sugar water for different $\Rey$, (a) $Re=125$, (b) $Re=600$, (c) $Re=1000$, (d) $Re=1400$, and (e) $Re=1700$. }
\label{fig:PLIF}
\end{figure}

The dynamics of a submerged liquid jet are governed by the competition among inertia, interfacial tension, viscous stresses, and buoyancy. The relative importance of these effects is quantified through the Reynolds number, $\Rey = \rho_d V_{\rm mean} D / \mu_d$, and the Weber number, $\We = \rho_c V_{\rm mean}^2 D / \sigma$, along with the density and viscosity ratios of the dispersed and continuous phases. In the present study, the Reynolds number was systematically varied over the range $\Rey = 125$-$1700$, spanning the laminar, transitional, and  turbulent regimes. The corresponding non-dimensional parameters for each experimental case are summarised in Table~\ref{table:ndn}, and representative instantaneous PLIF images are shown in figure~\ref{fig:PLIF}.
 
At $\Rey = 125$, the jet exits the nozzle in a laminar state and undergoes acceleration under the action of buoyancy, as evident in figure~\ref{fig:PLIF}(a). The dominance of buoyancy over inertia produces a pronounced contraction of the jet column, which ultimately disintegrates via the classical Rayleigh-Plateau (RP) instability. As $\Rey$ increases to 600, the jet enters a transitional regime in which neither laminar nor fully turbulent characteristics prevail. The jet column exhibits minimal radial contraction relative to the $\Rey = 125$ case and disintegrates primarily through the action of sinuous (lateral) instabilities, as evident in figure~\ref{fig:PLIF}(b). This sinuous mode reflects the emergence of asymmetric deformation of the jet column, marking a departure from the axisymmetric RP mechanism that governs breakup at lower Reynolds numbers.
 
For $\Rey \geq 1000$, inertial forces become dominant over viscous and interfacial tension effects, rendering the jet shear layer increasingly susceptible to 
Kelvin-Helmholtz (KH) instability. In this  regime, interfacial perturbations amplify rapidly, roll up into coherent vortical structures, and cascade to progressively smaller scales, culminating in turbulent fragmentation and droplet formation, as shown in figures~\ref{fig:PLIF}(c)-(e). The progressive transition in the dominant instability mechanism, from sinuous modes at $\Rey = 600$ to KH-driven breakup at $\Rey = 1000$, 1400, and 1700, is clearly reflected in the PLIF images, where the jet morphology evolves from organised sinuous oscillations to a highly disordered, fragmented interface. To further characterise the breakup dynamics and identify the dominant coherent structures underlying these regimes, a proper orthogonal decomposition (POD) analysis is presented in the following section.

\subsection{Modal Decomposition and Identification of Coherent Structures}
\label{sec:pod_plif}
To extract the dominant spatio-temporal structures from the dispersed-phase scalar field, proper orthogonal decomposition (POD) was applied to the PLIF dataset. POD, also referred to as the Karhunen-Lo\`eve decomposition or Principal Component Analysis (PCA), is a data-driven technique that decomposes an ensemble of flow snapshots into a hierarchy of orthogonal spatial modes ranked by their energy content~\citep{lumley1967structure, sirovich1987turbulence}.
 
Given a collection of $m$ snapshots of the scalar concentration field,
\begin{equation}
    \mathbf{x}_1,\,\mathbf{x}_2,\,\dots,\,\mathbf{x}_m \in \mathbb{R}^n,
\end{equation}
sampled at a uniform interval $\Delta t$, the fluctuating component of each snapshot is obtained by subtracting the ensemble mean:
\begin{equation}
    \mathbf{x}'_k = \mathbf{x}_k - \bar{\mathbf{x}}, 
    \qquad 
    \bar{\mathbf{x}} = \frac{1}{m}\sum_{k=1}^{m}\mathbf{x}_k.
\end{equation}
The fluctuation snapshots are assembled into the data matrix
\begin{equation}
    \mathbf{X}' = \bigl[\,\mathbf{x}'_1 \;\; \mathbf{x}'_2 \;\; \dots \;\; 
    \mathbf{x}'_m\,\bigr] \in \mathbb{R}^{n \times m}.
\end{equation}
Following the method of snapshots~\citep{sirovich1987turbulence}, the POD modes are 
obtained by solving the eigenvalue problem for the temporal correlation matrix 
$\mathbf{C} \in \mathbb{R}^{m \times m}$:
\begin{equation}
    \mathbf{C} = \frac{1}{m}\,\mathbf{X}'^{\!\top}\mathbf{X}', 
    \qquad 
    \mathbf{C}\,\mathbf{v}_j = \sigma_j^2\,\mathbf{v}_j,
\end{equation}
where the eigenvalues $\sigma_1^2 \geq \sigma_2^2 \geq \dots \geq \sigma_m^2 \geq 0$ are ordered by decreasing energy, and $\mathbf{v}_j \in \mathbb{R}^m$ are the corresponding eigenvectors. The $j$-th spatial POD mode is reconstructed in physical space as
\begin{equation}
    \boldsymbol{\phi}_j = \frac{1}{\sqrt{m}\,\sigma_j}\,\mathbf{X}'\,\mathbf{v}_j,
\end{equation}
with the modes orthonormal by construction, $\langle \boldsymbol{\phi}_i, 
\boldsymbol{\phi}_j \rangle = \delta_{ij}$. The relative energy contribution of mode $j$ is
\begin{equation}
    E_j = \frac{\sigma_j^2}{\displaystyle\sum_{k=1}^{m}\sigma_k^2}.
    \label{eq:pod_energy}
\end{equation}
In practice, retaining only the leading $r \ll m$ modes provides a low-rank approximation that suppresses low-energy noise while faithfully representing the dominant coherent structures.
\begin{figure}
\centering
\includegraphics[width=0.85\textwidth]{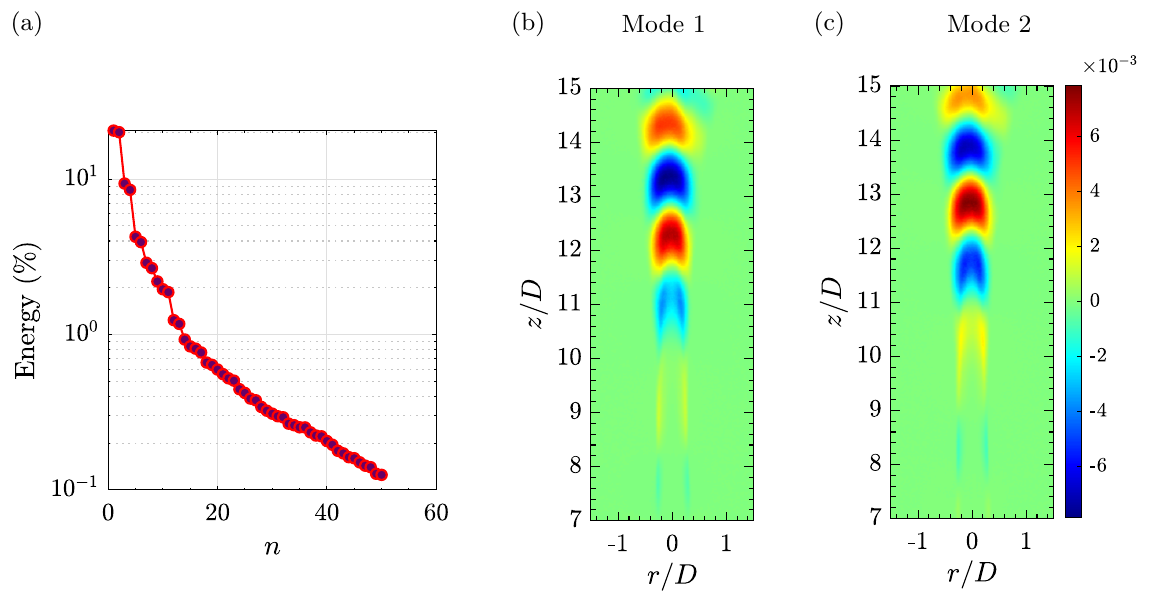}
\caption{(a) Relative energy distribution across POD modes for $Re = 125$. Spatial contour plots of the reconstructed POD modes for $Re = 125$: (b) Mode 1 and (c) Mode 2. }
\label{fig:pod125}
\end{figure}
 
\begin{figure}
\centering
\includegraphics[width=0.85\textwidth]{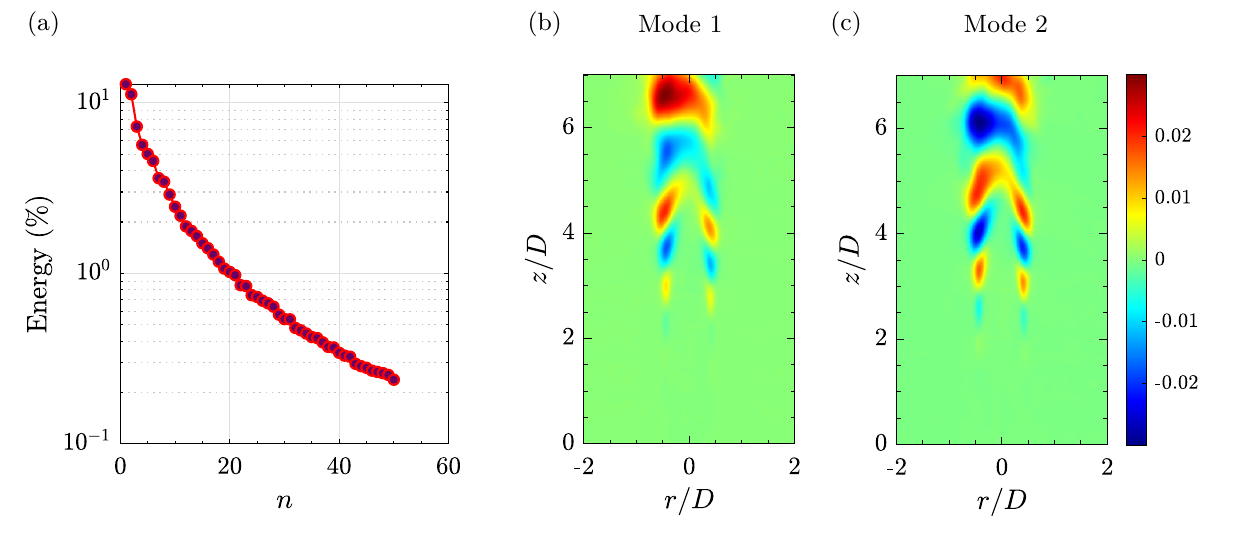}
\caption{(a) Relative energy distribution across POD modes for $Re = 600$. Spatial contour plots of the reconstructed POD modes for $Re = 600$: (b) Mode 1 and (c) Mode 2. }
\label{fig:pod600}
\end{figure}
 
\begin{figure}
\centering
\includegraphics[width=0.85\textwidth]{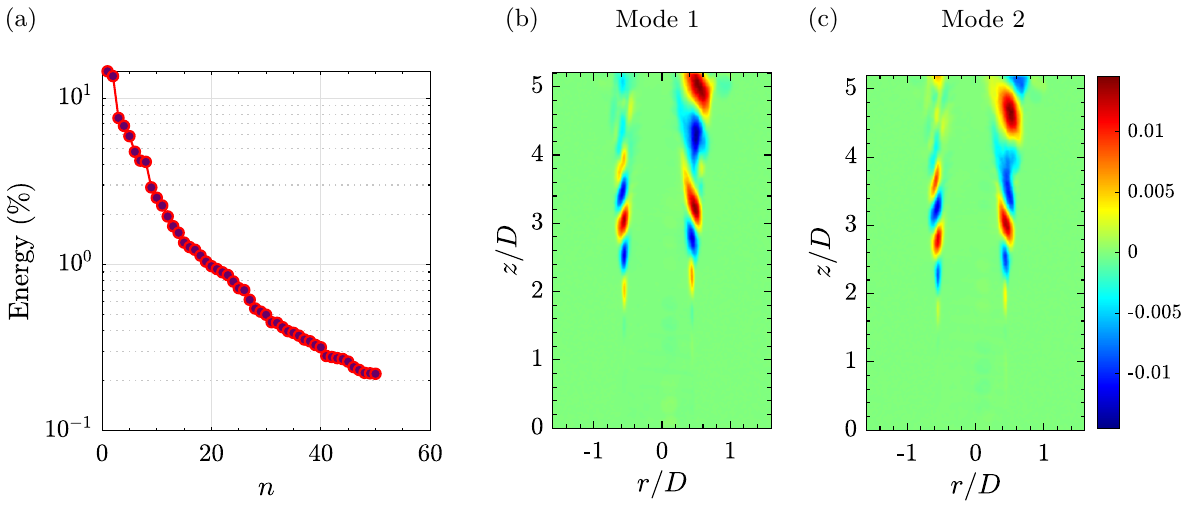}
\caption{(a) Relative energy distribution across POD modes for $Re = 1000$. Spatial contour plots of the reconstructed POD modes for $Re = 1000$: (b) Mode 1 and (c) Mode 2. }
\label{fig:pod1000}
\end{figure}
The POD analysis was performed on $N_s = 2000$ PLIF snapshots for $\Rey = 125$, $600$, and $1000$. Prior to decomposition, the ensemble mean was subtracted from each snapshot so that the resulting modes reflect only the dynamic, fluctuating content of the scalar field. The relative energy spectra $E_j$ for the three Reynolds numbers are shown in figures~\ref{fig:pod125}(a), \ref{fig:pod600}(a), and \ref{fig:pod1000}(a). In all cases, the energy decays rapidly beyond the first few modes, confirming that the large-scale coherent structures are well represented by the leading POD modes alone.
 
At $\Rey = 125$, the two dominant modes (figures~\ref{fig:pod125} (b,c)) exhibit a varicose (axisymmetric) spatial structure, consistent with the  Rayleigh-Plateau-like instability that governs breakup in the low-inertia regime. Modes 1 and 2 share the same underlying instability but are spatially phase-shifted relative to one another, a signature of the travelling-wave character of the Rayleigh-Plateau (RP) mechanism.
 
At $\Rey = 600$, the axial symmetry of the varicose mode is broken and a sinuous (lateral) wave pattern emerges as the energetically dominant structure 
(figure~\ref{fig:pod600}b), signalling a transition from purely axisymmetric to asymmetric deformation of the jet column. Mode 2 (figure~\ref{fig:pod600}c) retains a residual varicose signature, and the simultaneous presence of both sinuous and varicose modes confirms that $\Rey = 600$ occupies a transitional regime in which the two instability mechanisms coexist and compete near the breakup region.
 
At $\Rey = 1000$, representative of a  turbulent jet, the leading POD modes (figures~\ref{fig:pod1000} (b,c)) are governed by Kelvin-Helmholtz-like instability. The jet core does not exhibit any organised structures, whereas the surrounding shear layer experiences Kelvin-Helmholtz (KH)-induced roll-up, generating the energetically dominant features represented by the leading modes. The progressive shift in modal character from axisymmetric RP modes at $\Rey = 125$, through coexisting sinuous and varicose structures at $\Rey = 600$, to KH-type shear-layer roll-up at $\Rey = 1000$, provides direct evidence of the instability transition documented in the PLIF images of figure~\ref{fig:PLIF}. Note that the POD modes are interpreted as signatures of the dominant flow structures rather than direct measurements of instability growth rates or wavelengths.
 
Having established the dominant instability mechanisms through POD analysis, we now turn to the quantitative characterisation of the mean flow and turbulence statistics using PIV measurements. For the turbulent cases, turbulent velocity fluctuations continuously deform droplet surfaces, and the resulting droplet size distribution reflects the dynamic competition between turbulent breakup, which fragments droplets through the action of turbulent stresses, and coalescence, which promotes droplet growth through binary collisions in regions of elevated dispersed-phase concentration. The mean flow evolution, turbulent statistics, and droplet size distributions under these conditions are analysed in the sections that follow.

\subsection{Analysis of Turbulent Characteristics}
\label{sec:turbulent_characteristics}
\subsubsection{Mean Flow Evolution}
\begin{figure}
\centering
\includegraphics[width=0.99\textwidth]{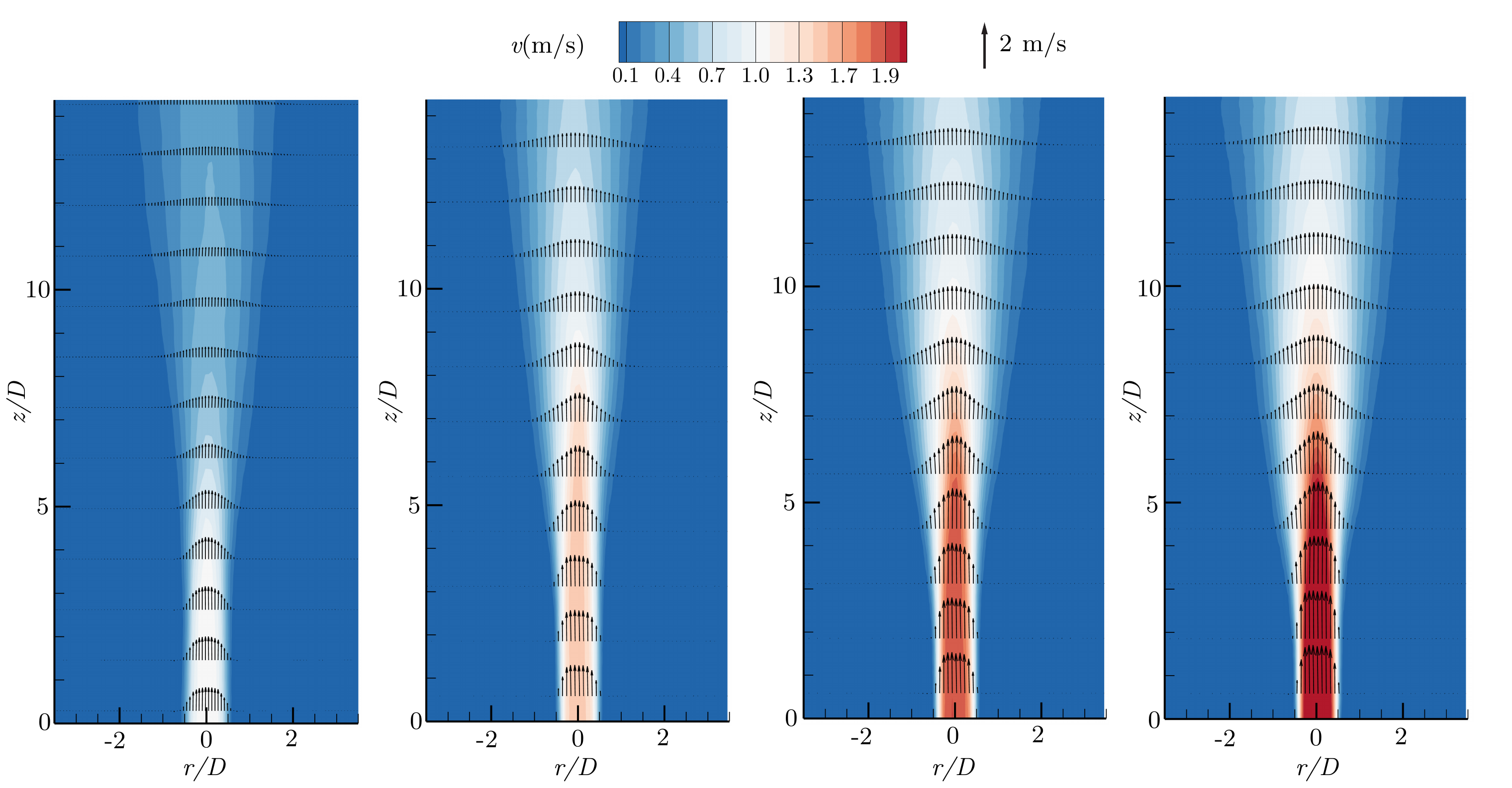}
\caption{Average velocity contours superimposed with velocity vectors for different $\Rey$ = (a) 600, (b) 1000, (c) 1400, and (d) 1700. }
\label{fig:vel_contours}
\end{figure}

\begin{figure}
\centering
\includegraphics[scale=0.6]{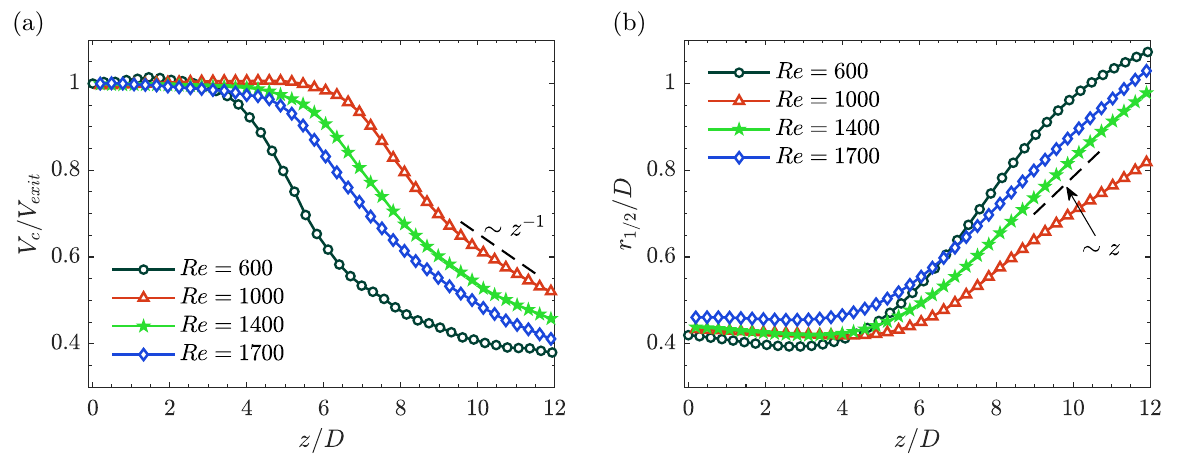}
\caption{(a) centreline velocity ($V_{c}$) normalised by the exit centreline velocity ($V_{exit}$) and, 
(b) Velocity half-width ($r_{1/2}$) normalised by the nozzle diameter ($D$), for different $\Rey$.}
\label{fig:centerline_vel}
\end{figure}

Figure~\ref{fig:vel_contours} shows the mean velocity contours, superimposed with velocity vectors, for different Reynolds numbers. To quantify the streamwise evolution of the jet and its lateral spreading, the normalised centreline velocity $V_c/V_{\rm exit}$ and the velocity half-width $r_{1/2}$ are plotted as functions of axial distance in figure~\ref{fig:centerline_vel}, where $r_{1/2}$ denotes the radial location at which the mean velocity falls to half the centreline value and all positions are non-dimensionalised by $D$.
 
The centreline velocity evolution shown in figures~\ref{fig:vel_contours} and \ref{fig:centerline_vel}(a) reveals a clear dependence on Reynolds number. At $\Rey = 600$, buoyancy accelerates the jet immediately downstream of the nozzle, producing a local rise in $V_c$, before viscous drag and jet disintegration cause a rapid decline. At higher Reynolds numbers, exit inertia dominates buoyancy and $V_c$ remains approximately constant throughout the intact jet column, before declining sharply at the breakup point. Beyond breakup, $V_c \sim z^{-1}$ for all high-$\Rey$ cases, consistent with  turbulent jet scaling. The breakup length exhibits a non-monotonic dependence on Reynolds number, increasing from $\Rey = 600$ to $1000$ before decreasing monotonically at higher $\Rey$. This behaviour reflects a transition in the dominant disintegration mechanism: in the transitional regime, inertia stabilises the jet column against interfacial perturbations and lengthens the breakup length, whereas in the atomisation regime, the intensification of Kelvin-Helmholtz instabilities and turbulent fluctuations progressively overcomes this stabilising effect, consistent with the POD analysis of Section~\ref{sec:pod_plif}.

The evolution of the velocity half-width, shown in figure~\ref{fig:centerline_vel}(b), further distinguishes the transitional and atomisation regimes. In the atomisation regime ($\Rey \geq 1000$), $r_{1/2}$ remains nearly constant over the intact jet column and increases at a broadly similar rate across all cases once the jet has disintegrated. The slightly elevated initial value of $r_{1/2}$ at $\Rey = 1700$ relative to $\Rey = 1000$ and $1400$ reflects enhanced inertia-driven momentum diffusion in the near-nozzle region prior to breakup. In the transitional regime ($\Rey = 600$), by contrast, $r_{1/2}$ exhibits a slight initial decrease attributable to jet thinning under buoyancy-driven acceleration, which reduces the jet cross-section by mass conservation. Following breakup, $r_{1/2}$ increases at a markedly higher rate than in the atomisation cases, indicating that lateral momentum transport is significantly more vigorous in the transitional regime. 
 
\begin{figure}
\centering
\includegraphics[scale=0.45]{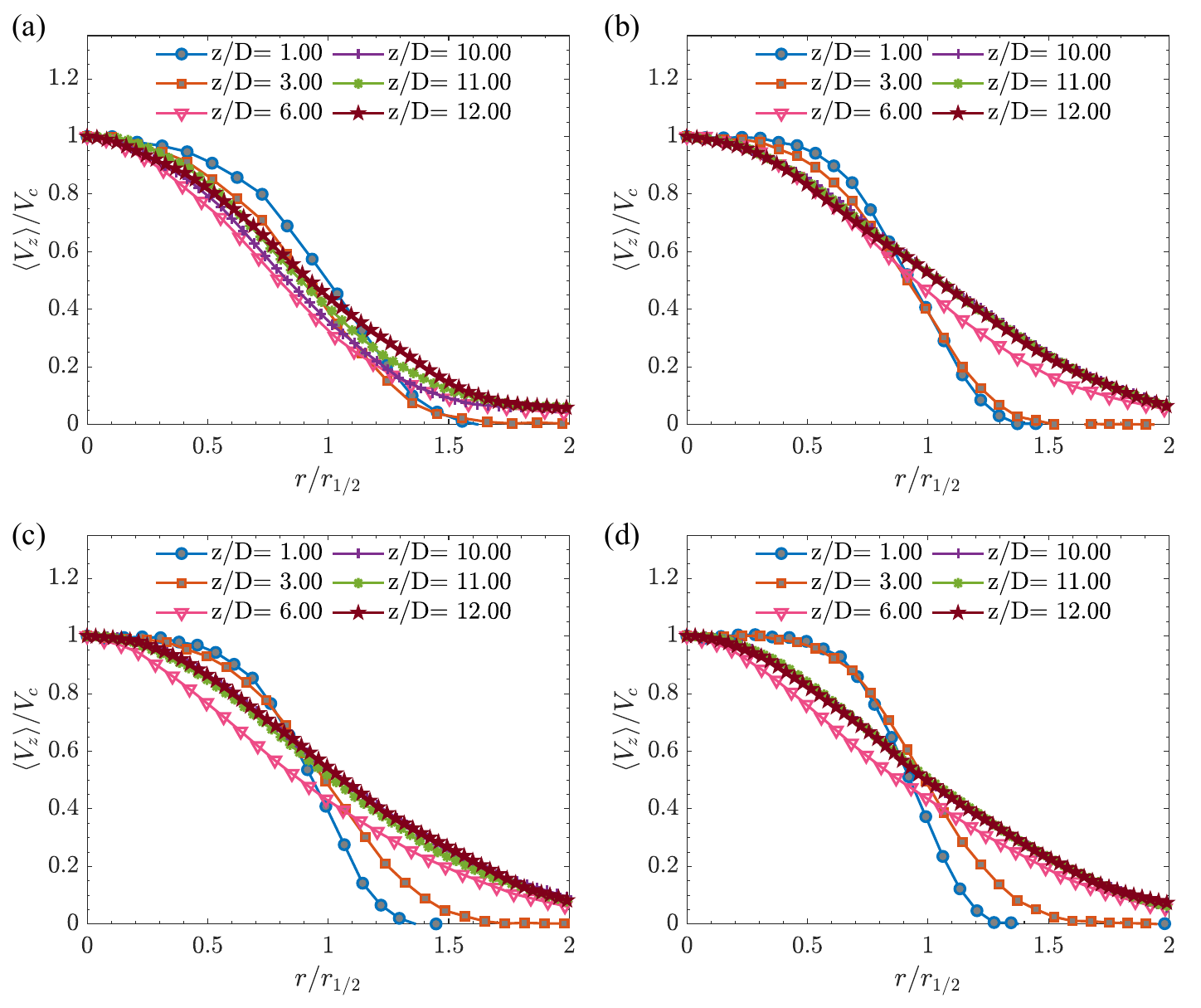}
\caption{Axial velocity profiles, scaled by the centreline velocity and velocity half-width at different axial locations, for (a) $Re = 600$, (b) $Re = 1000$, (c) $Re = 1400$, and (d) $Re = 1700$.}
\label{fig:vel_profiles}
\end{figure}

The self-similar character of the velocity field is examined in figure~\ref{fig:vel_profiles}, where axial velocity profiles at successive downstream stations are normalised by the local centreline velocity and plotted against $r/r_{1/2}$. For the atomisation-regime cases ($Re>600$), the profiles collapse onto a common Gaussian-like curve for $z/D \gtrsim 6$, confirming the onset of self-similarity downstream of the breakup point. Upstream of breakup, the profiles retain their initial top-hat shape and deviate from self-similarity due to momentum diffusion within the dispersed phase and the growth of interfacial instabilities. Taken together, these results demonstrate that the downstream evolution of the mean axial velocity field is fully characterised by the two integral length and velocity scales $r_{1/2}(z)$ and $V_c(z)$, which form the basis of the similarity analysis adopted in the population balance model presented in Section~\ref{sec:pbm}.

\subsubsection{Turbulent Quantities and Reynolds Stresses}
\label{sec:reynolds_stresses}

The velocity fluctuations are defined as $u'=u-\overline{u}$ and $v'=v-\overline{v}$, and the Reynolds shear stress is expressed as the kinematic correlation $\langle u'v'\rangle$. The corresponding momentum flux would be $\rho\langle u'v'\rangle$ \citep{amielh1996velocity}. Since the density ratio of the two phases is relatively small ($\rho_c/\rho_d\approx1.20$), the density weighting is neglected here, and $\langle u'v'\rangle$ is used to characterise the turbulent structure.
 
\begin{figure}
\centering
\includegraphics[scale=0.45]{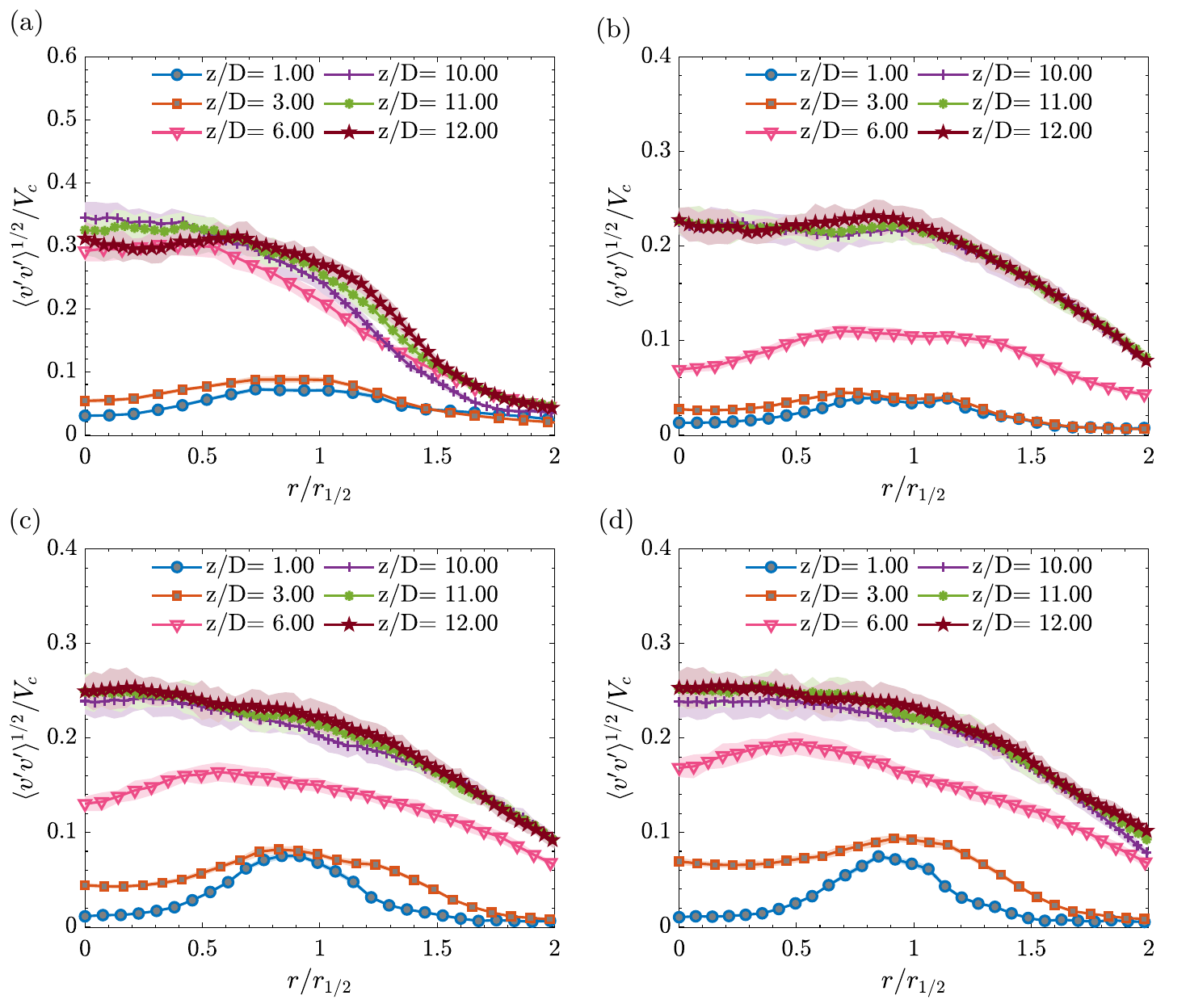}
\caption{Axial RMS velocity profiles, normalised by the centreline velocity and jet half-width at various axial locations, for (a) $Re = 600$, (b) $Re = 1000$, (c) $Re = 1400$, and (d) $Re = 1700$. The shaded regions represent the estimated uncertainty associated with the convergence bias.}  
\label{fig:vrms}
\end{figure}

\begin{figure}
\centering
\includegraphics[scale=0.45]{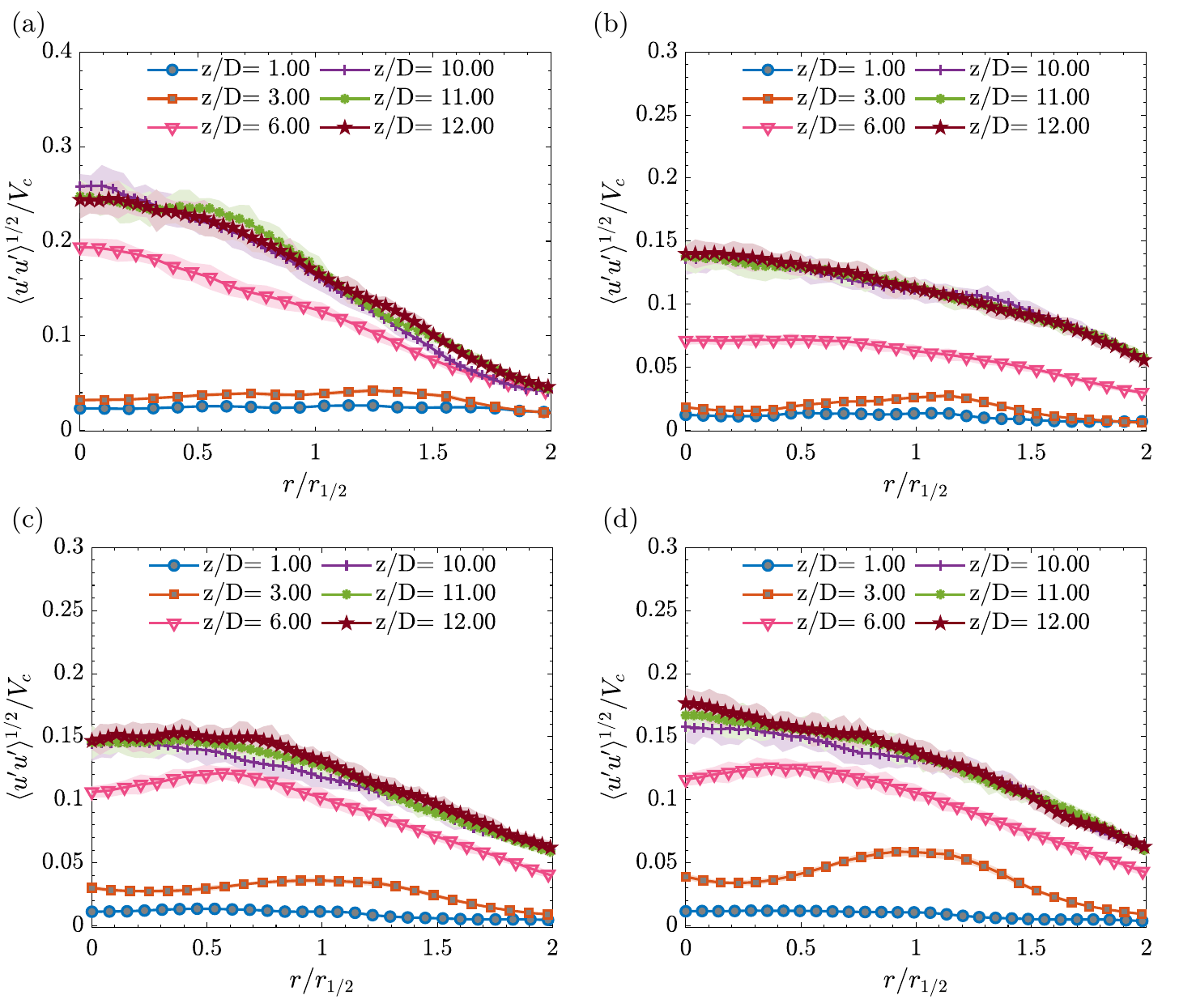}
\caption{Radial RMS velocity profiles, normalised by the centreline velocity and jet half-width at various axial locations, for (a) $Re = 600$, (b) $Re = 1000$, (c) $Re = 1400$, and (d) $Re = 1700$. The shaded regions represent the estimated uncertainty associated with the convergence bias.} 
\label{fig:urms}
\end{figure}

Figures~\ref{fig:vrms} and \ref{fig:urms} present the axial and radial velocity fluctuations, $\langle v'v' \rangle^{1/2}$ and $\langle u'u' \rangle^{1/2}$, 
normalised by the local centreline velocity, at different axial distances from the nozzle. The radial coordinate is non-dimensionalised by $r_{1/2}$.
 
Before breakup, both fluctuation components (figures~\ref{fig:vrms},\ref{fig:urms}) attain their peak near the jet radius, a characteristic of  KH-driven shear-layer instability arising from the velocity difference between the two phases. The fluctuation intensity grows monotonically with downstream distance as the instabilities amplify, culminating in the disintegration of the jet into droplets and ligaments. Beyond the breakup point ($z/D\gtrsim 10$), both components collapse onto self-similar profiles when scaled by $V_c$ and $r_{1/2}$, with the peak shifting to the centreline in accordance with the decaying mean velocity field. This self-similar behaviour confirms that the turbulent fluctuation field is governed by the same integral scales that characterise the mean flow evolution.
\begin{figure}
\centering
\includegraphics[scale=0.45]{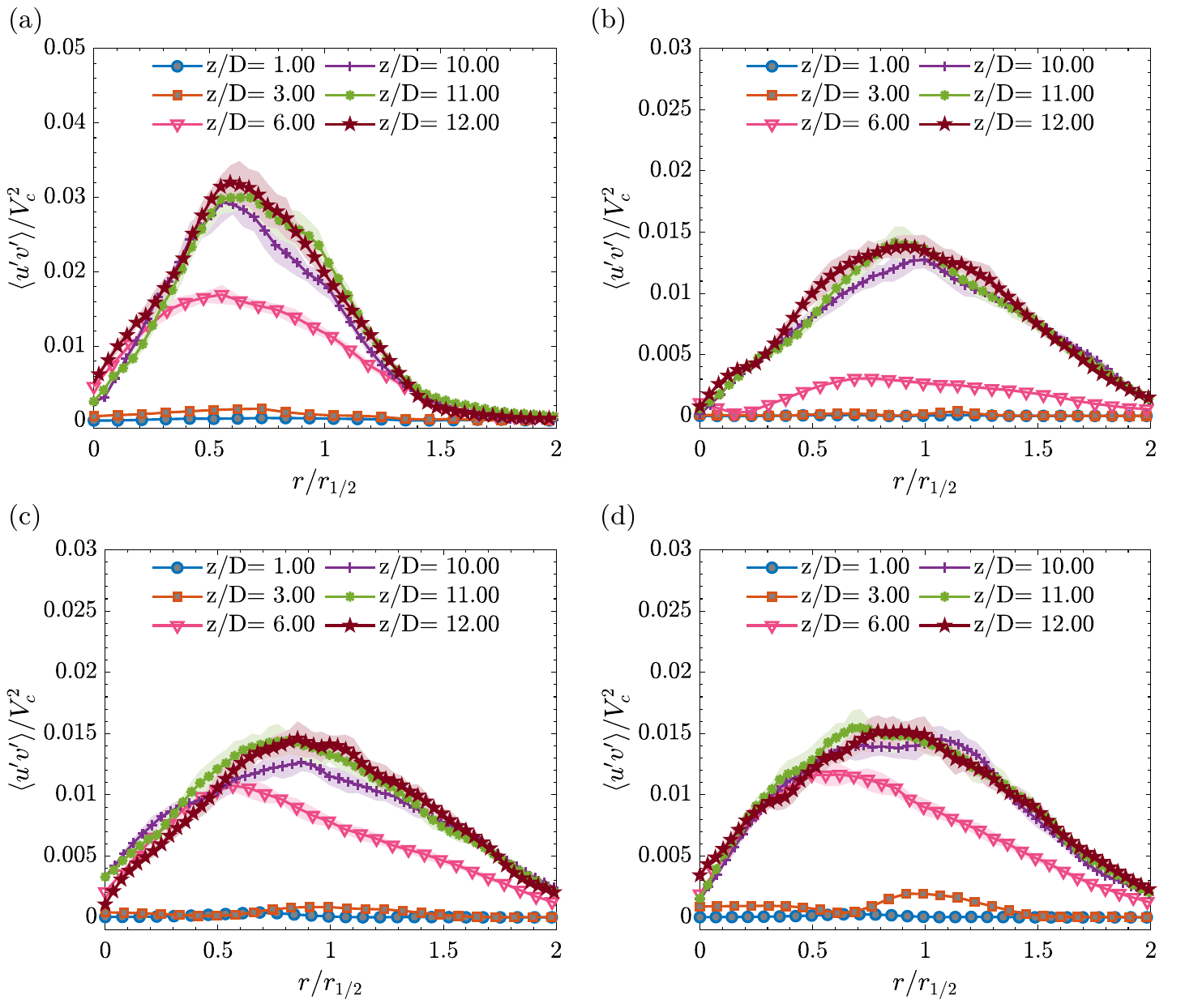}
\caption{Profiles of Reynolds stresses, scaled by the centreline velocity and velocity half-width, at different axial locations, for (a) $Re = 600$, (b) $Re = 1000$, (c) $Re = 1400$, and (d) $Re = 1700$. The shaded regions represent the estimated uncertainty associated with the convergence bias.} 
\label{fig:re_stress}
\end{figure}

The Reynolds shear stress $\langle u'v' \rangle$, normalised by $V_c^2$ and plotted against $r/r_{1/2}$ in figure~\ref{fig:re_stress}, exhibits an analogous spatial structure. The stress is minimum at the centreline and peaks in the shear layer, confirming that interfacial instabilities and droplet production are concentrated at the jet periphery. As the instabilities grow downstream, the Reynolds stresses intensify and ultimately drive jet disintegration, beyond 
which they collapse onto self-similar profiles when scaled by $V_c$ and $r_{1/2}$. The linear growth of $r_{1/2}$ with axial distance captures the progressive lateral spreading of the dispersed phase in the post-breakup region, reinforcing its role as the appropriate integral length scale for characterising jet spreading.
Note that Reynolds stress profiles collapse at approximately $z/D \gtrsim 10$, whereas the mean velocity profiles become self-similar earlier, at approximately $z/D \gtrsim 6$. This is expected, as higher-order turbulence statistics generally require a longer development length than the mean flow to attain self-similarity.

\subsection{Phase-Conditioned Mean Velocity and Reynolds Stress Profiles}
\label{sec:phase_conditioned}

The combined mean velocity and Reynolds stress statistics presented in Section~\ref{sec:turbulent_characteristics} represent an average over both the dispersed oil phase and the continuous sugar-water phase at each measurement location. Since the PLIF measurements provide the instantaneous spatial distribution of the dispersed phase simultaneously with the PIV velocity field, this information can be used to condition the velocity statistics on the locally dominant phase, following the approach of \citet{burridge2017conditional} and \citet{xue2021flow}. This phase-conditioned analysis isolates the individual contribution of each phase to the combined statistics reported in Section~\ref{sec:turbulent_characteristics}, and provides additional insight into the momentum and turbulence exchange between the dispersed and continuous phases as the jet evolves downstream.

\subsubsection{Phase indicator function and conditional averaging}

At each time instant and spatial location, a phase indicator function is constructed from the instantaneous PLIF volume fraction field,
\begin{equation}
    \Phi(r,z,t) =
    \begin{cases}
        1, & \text{oil present at } (r,z,t), \\
        0, & \text{water present at } (r,z,t), 
    \end{cases}
    \label{eq:heaviside}
\end{equation}
using the CNN-based segmentation described in Section~\ref{sec:dsd}. The phase-conditioned ensemble-averaged velocity in the oil, $V_{i,o}(r,z)$, and in the water, $V_{i,w}(r,z)$, are then defined as
\begin{equation}
    V_{i,o}(r,z) = \frac{1}{\langle \Phi\rangle N}\sum_{k=1}^{N}
    \Phi(r,z,t_k)\,v_i(r,z,t_k),
    \label{eq:Uio}
\end{equation}
\begin{equation}
    V_{i,w}(r,z) = \frac{1}{N - \langle \Phi\rangle N}\sum_{k=1}^{N}
    \left[1-\Phi(r,z,t_k)\right] v_i(r,z,t_k),
    \label{eq:Uiw}
\end{equation}
where $N$ is the total number of instantaneous realisations at a given
location and $\langle \Phi \rangle$ is the local time-averaged oil fraction. Here, subscript $i$ and $j$ represents the axial and radial velocity components, respectively. The corresponding phase-conditioned Reynolds stresses are obtained analogously, using the fluctuating velocity $v_i' = v_i - V_{i,n}$
($n = o, w$) about each phase-specific mean,
\begin{equation}
    \left\langle v_{i,o}'v_{j,o}'\right\rangle =
    \frac{1}{\langle \Phi\rangle N}\sum_{k=1}^{N}
    \Phi(r,z,t_k)\,v_i'(r,z,t_k)\,v_j'(r,z,t_k),
    \label{eq:RSo}
\end{equation}
\begin{equation}
    \left\langle v_{i,w}'v_{j,w}'\right\rangle =
    \frac{1}{N-\langle \Phi\rangle N}\sum_{k=1}^{N}
    \left[1-\Phi(r,z,t_k)\right] v_i'(r,z,t_k)\,v_j'(r,z,t_k).
    \label{eq:RSw}
\end{equation}

\begin{figure}
\centering
\includegraphics[scale=0.65]{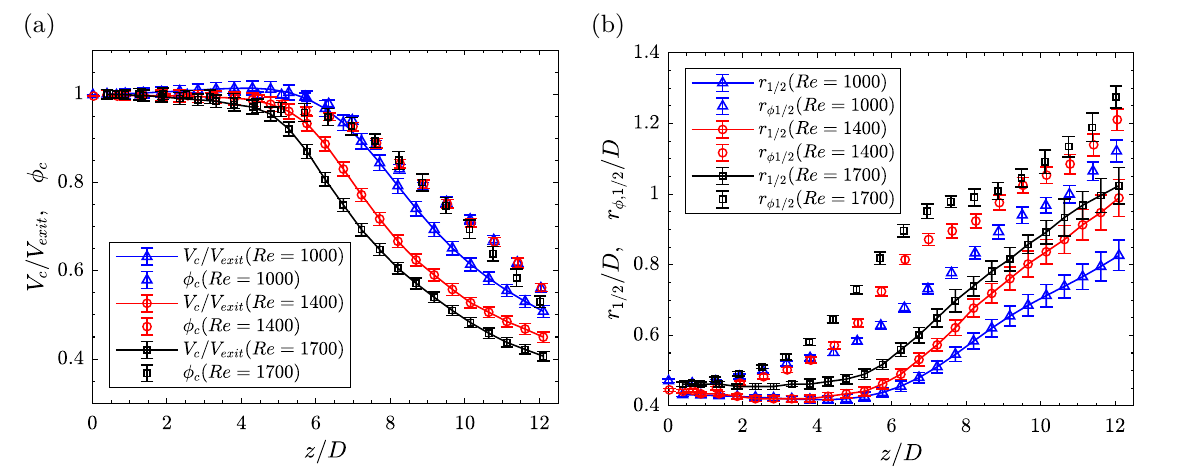}
\caption{Comparison between PIV and PLIF measurements of (a) centreline velocity ($V_{c}$), normalised by the exit centreline velocity ($V_{exit}$) (symbols with solid lines and error bars), centerline volume fraction(symbols with error bars) and (b) velocity half-width ($r_{1/2}$),normalised by the nozzle diameter ($D$) (symbols with solid lines and error bars), oil fraction half width($r_{\phi, 1/2}$)(symbols with error bars) for different $\Rey$.}
\label{fig:vol_fraction}
\end{figure}

\subsubsection{Evolution of centreline volume fraction and volume fraction-based half-width}
In this section, we examine the axial evolution of the mean centreline volume fraction, $ \phi_{c} (z) = \langle \Phi(0,z,t)\rangle$, and the volume-fraction-based half-width, $r_{\phi,1/2}$. Figure~\ref{fig:vol_fraction}(a) compares $ \phi_{c} $ with the mean centreline velocity $V_c$. Although both quantities decrease downstream, their decay is governed by fundamentally different physical mechanisms. The centreline velocity decreases as momentum is progressively transferred to the surrounding fluid through shear at the interface and momentum diffusion. This process begins immediately downstream of the potential core and results in a gradual reduction of $V_c$, well before significant interfacial breakup occurs. In contrast, the centreline volume fraction is governed primarily by the evolution of the liquid-liquid interface. As long as the jet core is intact, the centreline lies entirely within the oil phase, so $\phi_{c} = 1$, and it only begins to fall once pinch-off sets in. This is exactly what is observed: $\phi_{c}$ remains at unity over a finite distance for all $Re$, marking the extent of the intact jet core, then decreases monotonically once breakup begins, with a steeper decay at higher $Re$, indicating faster dispersion.

The contrasting physical mechanisms governing velocity and volume fraction are also reflected in the evolution of their respective half-widths, shown in Figure~\ref{fig:vol_fraction}(b). Here, $r_{\phi,1/2}$ denotes the radial location at which the volume fraction decreases to $\phi_{c}/2$.  The velocity half-width grows at an approximately constant rate from the potential-core region onward, since momentum diffuses continuously into the surrounding fluid irrespective of interfacial morphology. On the other hand, $r_{\phi,1/2}$ tracks the dispersed phase directly and therefore depends on breakup rather than diffusion. It remains nearly constant through the intact core, increases sharply at the onset of primary breakup as the jet fragments into ligaments and droplets that spread radially.
Farther downstream, after the breakup process is well established, $r_{\phi,1/2}$ exhibits a slower, approximately linear growth, characteristic of turbulent dispersion. The distinct evolution of the velocity and volume-fraction widths highlights the difference between momentum transport and phase dispersion and provides the motivation for the phase-conditioned analysis presented in the following section, where the contributions of the individual phases to the turbulent statistics are examined at different axial locations.

\subsubsection{Evolution of phase-conditioned mean velocity and turbulence}

\begin{figure}
    \centering
    \includegraphics[scale=0.55]{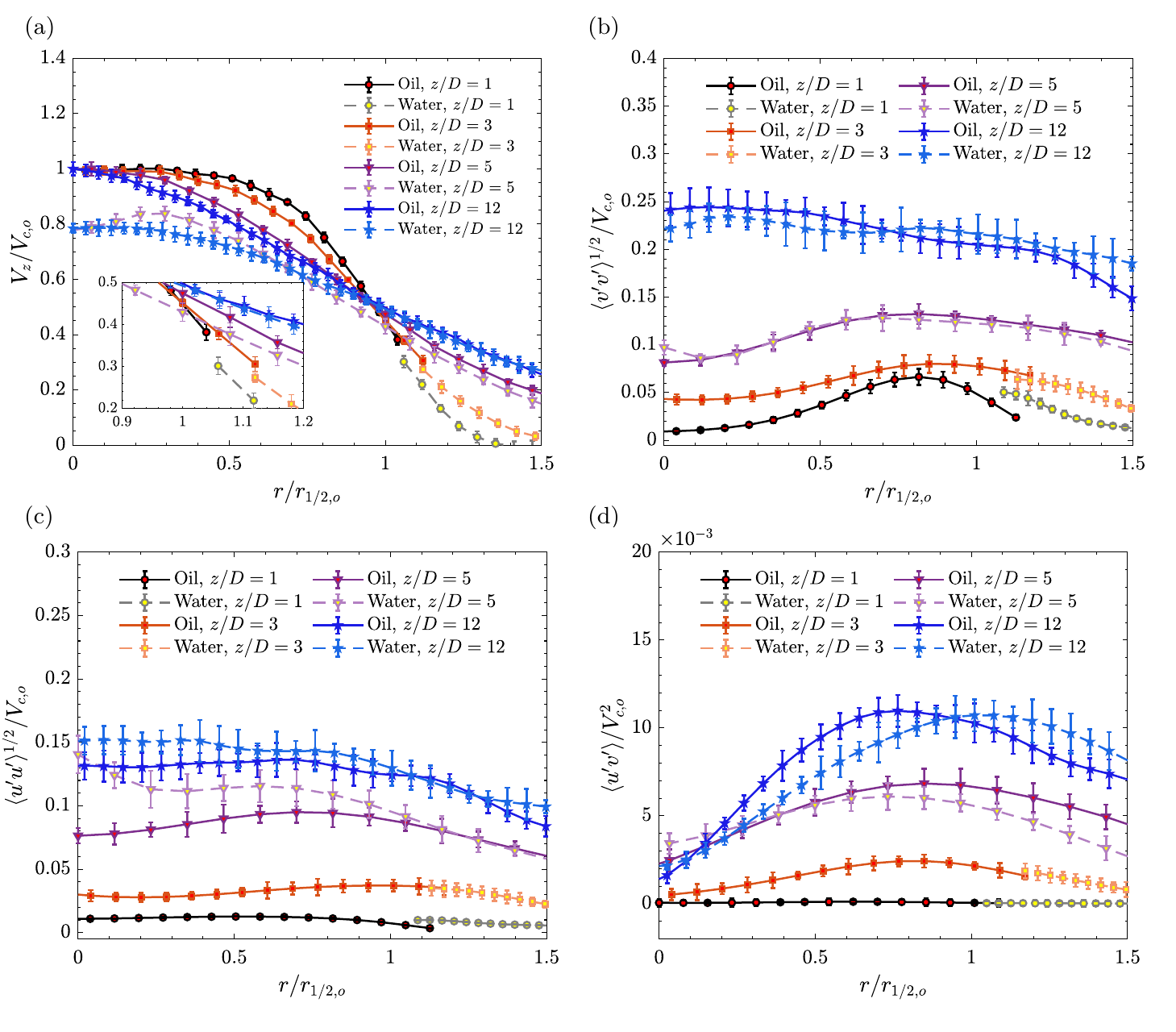}
    \caption{Phase-dependent (a) normalized mean axial velocity, (b) normalized axial velocity fluctuations ($\langle v'v' \rangle^{1/2}/V_{c,o}$), (c) normalized radial velocity fluctuations ($\langle u'u'\rangle^{1/2}/V_{c,o}$), and (d) normalized Reynolds shear stress ($\langle u'v'\rangle/V_{c,o}^2$). Here, $V_{c,o}$ and $r_{1/2,o}$ denote the mean centerline velocity and velocity half-width, respectively, of the oil phase. The error bars represent the estimated uncertainty associated with the convergence bias.}
    \label{fig:mean_RS_1400}
\end{figure}

Figure~\ref{fig:mean_RS_1400} presents the phase-conditioned mean axial
velocity and Reynolds stress profiles at $Re=1400$, at different axial locations, spanning the region from nozzle exit to the far field.  The evolution of mean axial velocity of two phases at different axial locations is illustrated in figure~\ref{fig:mean_RS_1400}(a). As is evident, near the nozzle ($z/D = 1$), the high-momentum oil jet induces drag on the surrounding fluid at the interface. Near the breakup region ($z/D = 5$), the oil-phase mean velocity $V_{z,o}$ exceeds the water-phase velocity $V_{z,w}$ close to the jet axis, as oil retains the majority of the injected momentum. The velocity difference between the phases is largest near the jet centreline, and decreases with radial distance, where entrained water and droplets are more thoroughly intermixed. Farther downstream, at $z/D = 12$, the difference between $V_{z,o}$ and $V_{z,w}$ progressively diminishes, indicating momentum equilibration between the phases as turbulent mixing redistributes momentum from the faster oil droplets to the surrounding continuous phase.

Figures~\ref{fig:mean_RS_1400}(b) and (c) show the axial and radial rms velocity fluctuations, and  figure~\ref{fig:mean_RS_1400}(d) the Reynolds shear stress, conditioned on the oil and water phases. Close to the nozzle exit ($z/D$=1), in the shear-layer region surrounding the jet core, both the rms components and the Reynolds stress are consistently higher in the water phase, reflecting the elevated turbulence generation associated with interfacial deformation. In the breakup region ($z/D=5$),  close to the centreline, the water-phase turbulence exceeds that of the oil, associated with more efficient turbulent mixing of the continuous phase and entrainment toward the centreline. With increasing axial distance, the difference between the phase-conditioned rms fluctuations and Reynolds stress diminishes, as the dispersed and continuous phases progressively homogenize downstream of the breakup region.  These results show that the turbulence statistics differ between the two phases near the breakup region, converging downstream as the dispersed and continuous phases mix and exchange momentum. The phase-conditioned statistics at the other $Re$ values, not presented here, also exhibit similar trends. 

With the mean flow evolution and turbulence characteristics established, we now turn to the droplet size distributions obtained from PLIF measurements.
 
\subsection{Droplet Size Distribution}
\label{sec:dsd}
\begin{figure}
\centering
\includegraphics[scale=0.70]{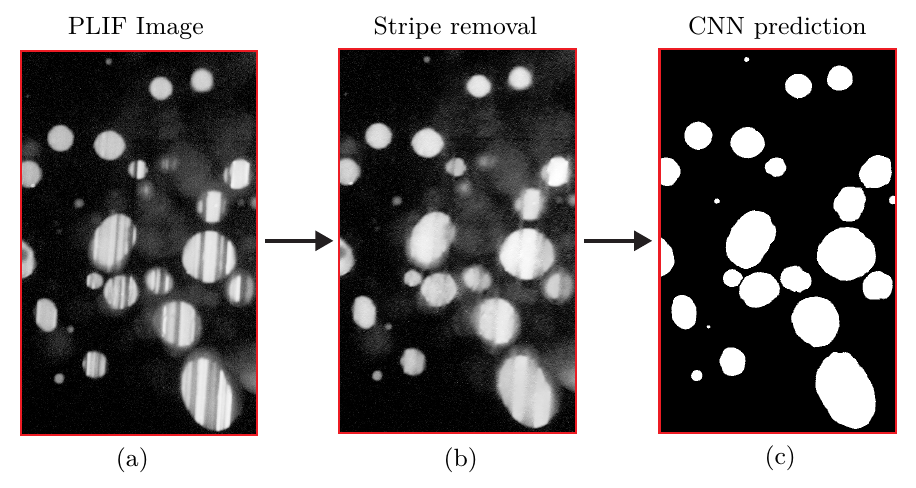}
\caption{PLIF image pre-processing steps for droplet size distribution analysis, (a) Raw PLIF image, (b) stripe-corrected image obtained using variational stripe-removal technique, and (c) the resulting binary image generated through CNN-based semantic segmentation.}
\label{fig:cnn_prediction}
\end{figure}
\subsubsection{Image Processing and Droplet Detection}
 
Droplet sizes are extracted from the PLIF images through a sequence of processing steps, illustrated in figure~\ref{fig:cnn_prediction}. The raw PLIF images contain stripe artefacts arising from laser refraction at multiple liquid interfaces; even a refractive index mismatch at the fourth decimal place is sufficient to introduce these distortions. 
They are removed prior to droplet detection using the variational stripe-removal method of \citet{Rottmayer25}, which separates each image into a stripe-free component and a directional stripe component and retains the former. Conventional intensity thresholding cannot reliably distinguish out-of-focus droplets and glare artifacts, which arise from multiple reflections of the laser sheet at the droplet and ligament interfaces.  
Droplet segmentation is therefore performed using a semantic segmentation approach based on the U-Net convolutional neural network architecture~\citep{ronneberger2015u}. The network was trained on 200 manually annotated images, supplemented by data augmentation including scaling, shearing, rotation, and elastic deformation to increase the effective training set size. 
The droplet size distribution is obtained from the binary images generated by the CNN using the built-in MATLAB function \texttt{regionprops}, which extracts the geometric properties of connected objects, including area, circularity, eccentricity, and equivalent circular diameter. The continuous jet is first excluded from the analysis based on an area constraint. Elongated and irregular liquid blobs are then excluded using geometric filtering criteria based on eccentricity and circularity. Eccentricity, defined as the ratio of the distance between the foci of the equivalent ellipse to its major-axis length, ranges from 0 for a perfect circle to 1 for a line; objects with an eccentricity greater than 0.8 are discarded to eliminate highly elongated structures. Circularity, defined as $C = 4\pi A/P^{2}$, where $A$ is the projected area and $P$ is the perimeter of the object, quantifies the deviation of an object from a perfect circle; objects with $C \leq 0.5$ are discarded to eliminate irregularly shaped blobs.  These filtering criteria effectively exclude elongated ligaments, residual jet fragments, and irregularly shaped blobs, providing a reliable
dataset for the subsequent droplet size distribution analysis.
 To ensure statistical independence, the residence time of droplets within the imaging window was estimated from the frame-to-frame cross-correlation of the droplet field. The droplets typically remained within the window for approximately 10 consecutive frames at 1000 Hz. Accordingly, the image sequence was sub-sampled at an interval exceeding this residence time, minimizing temporal correlation and avoiding repeated counting of the same droplet in the size statistics. The size distributions are computed from  1000 uncorrelated processed images. However, the tail of the distribution, corresponding to the largest droplets ($d/D \gtrsim 0.7$), shows a convergence bias, since large droplets represent a small fraction of the total droplet count in any given realisation. This convergence bias is represented as the error bars in the plots. A cutoff diameter of 2 pixels is used to define the minimum detectable droplet diameter. 
\subsection{Droplet distributions}
The droplet population is characterised through two statistical descriptors: number probability density $P_n(d)$ and volume probability density $P_v(d)$. The number probability density $P_n(d)$ represents the probability of finding a droplet of diameter $d$ in the population, normalised such that its 
integral over all diameters equals unity:
\begin{equation}
    P_n = \frac{n_i}{N \times \Delta d},
    \label{eq:num_prob}
\end{equation}
where $n_i$ is the number of droplets in the $i$th bin, $N$ is the total number of detected droplets, and $\Delta d$ is the bin width. The volume probability 
density $P_v(d)$ represents the fraction of the total droplet volume contributed by droplets of diameter $d$. Since droplet volume scales as $d^3$, $P_v$ weights larger droplets more heavily than $P_n$, and the two quantities are related by
\begin{equation}
    P_v(d) = \frac{d^3\, P_n(d)}{\displaystyle\int_0^{\infty} d^3\, P_n(d)\, \mathrm{d}d}.
    \label{eq:vol_prob}
\end{equation}
\begin{figure}
\centering
\includegraphics[scale=0.60]{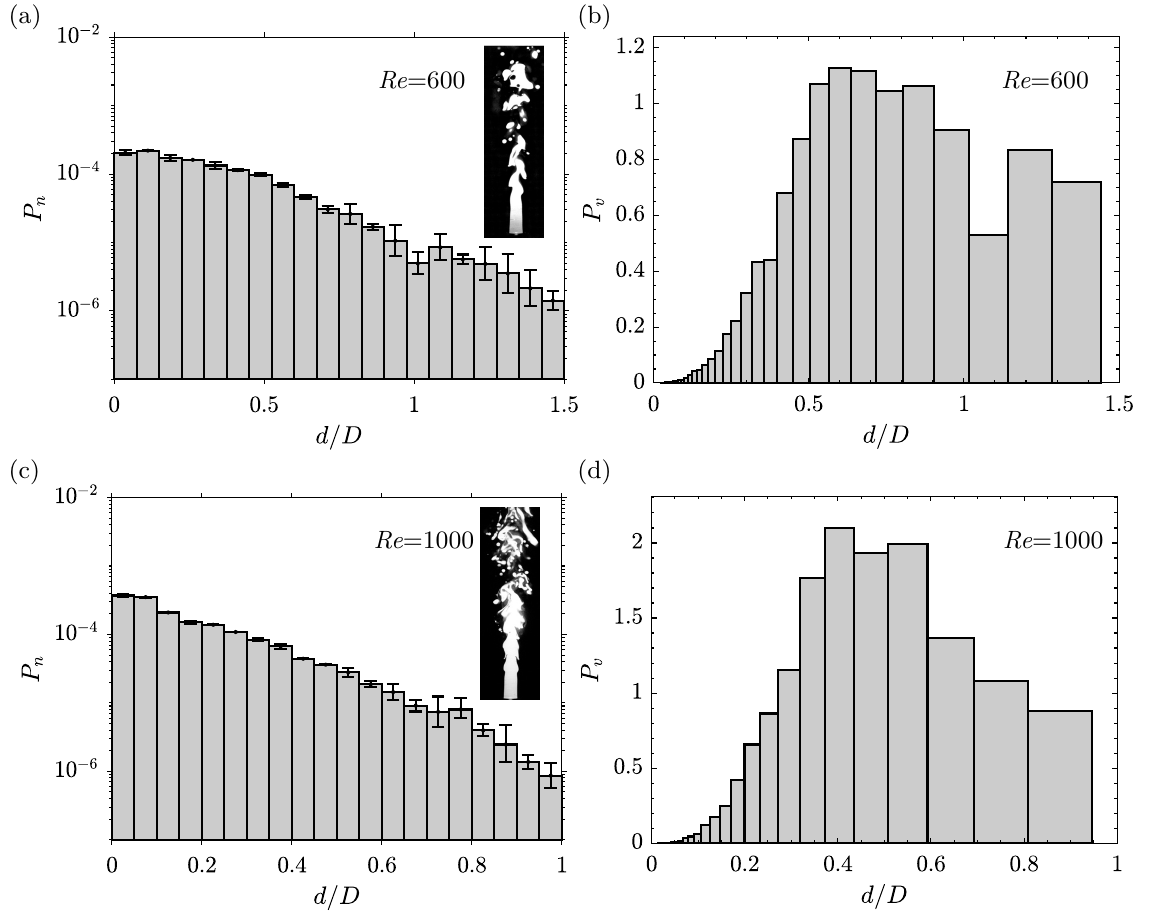}
\caption{Number probability density $P_n$ ((a), (c)) and volume probability density $P_v$ ((b), (d)) measured at $z/D = 20$ for $\text{Re} = 600$ ((a), (b)) and $\text{Re} = 1000$ ((c), (d)). Insets of (a) and (c) show the corresponding instantaneous PLIF images. The lower bound of the size distribution is 2 pixels ($\approx 100 \mu $m) and bin width is 500$\mu $m.}
\label{fig:pn_pv_transient}
\end{figure}
Both distributions are shown in figure~\ref{fig:pn_pv_transient} for $\Rey = 600$ and $1000$ at $z/D = 20$. In both cases, $P_n(d)$ decays monotonically with increasing diameter, confirming that smaller droplets are numerically dominant. However, the distribution at $\Rey = 600$ extends to larger diameters ($d/D = 1.5$), reflecting a greater proportion of large droplets compared to the $\Rey = 1000$ case.
 
A clearer distinction between the two regimes emerges from the volume probability density $P_v(d)$. At $\Rey = 600$, the distribution is distinctly bimodal 
(figure~\ref{fig:pn_pv_transient}b): the primary peak at small diameters corresponds to droplets produced by shear-induced breakup in the jet shear layer, while the secondary peak at larger diameters reflects the presence of large droplets formed through RP instability of the liquid column, as identified in Section~\ref{sec:pod_plif}, which remain largely unaffected by the weak turbulent stresses in the transitional regime. At $\Rey = 1000$, the volume distribution collapses to a single dominant mode (figure~\ref{fig:pn_pv_transient}d), indicating that droplet formation is governed entirely by shear breakup and turbulent stresses. The disappearance of the secondary peak with increasing Reynolds number thus reflects the growing dominance of inertial and turbulent effects over interfacial instabilities in controlling the droplet size distribution.
 
To complement the experimental measurements and provide a predictive framework for droplet size evolution, a one-dimensional (1D) ordinary differential equation (ODE) model is presented for large $Re$ cases following the population balance formulation of \citet{aiyer2020coupled}. The model exploits the self-similar behaviour of the mean and turbulent flow fields established in Section~\ref{sec:turbulent_characteristics} to reduce the axisymmetric population balance equation to a tractable 1D evolution equation along the jet centreline. In the present work, the original formulation is extended to incorporate droplet coalescence into the source term, and the model coefficients are determined directly from the experimental measurements. The formulation and numerical implementation are described in Section~\ref{sec:pbm}.

\section{ Population Balance Model}
\label{sec:pbm}
 
To predict the axial evolution of the droplet size distribution along the jet centreline, a one-dimensional (1D) ordinary differential equation (ODE) model is formulated following the population balance framework of \citet{aiyer2020coupled}. The formulation is extended to include droplet coalescence in the source term, and the model coefficients are obtained directly from the current experiments. The model incorporates the combined effects of axial advection, radial turbulent transport, droplet breakup, and droplet coalescence, and exploits the self-similar flow structure established in Section~\ref{sec:turbulent_characteristics} to reduce the governing equations to 1D form.
 
\subsection*{Model Formulation.}
 
The mean flow field is described within a Reynolds-averaged Navier-Stokes (RANS) framework closed with an eddy-viscosity model. The dispersed oil phase is 
represented using a sectional population balance approach, in which the droplet size spectrum is discretised into $M$ diameter classes. The mean volumetric 
concentration of the dispersed phase is expressed as
\begin{equation}
    \tilde{c}(z,r) = \sum_{k=1}^{M} \tilde{n}_k(z,r)\,\Omega_k,
    \label{eq:total_concentration}
\end{equation}
where $\tilde{n}_k$ is the number density of droplets of diameter $D_k$ per unit volume of the fluid, and $\Omega_k = \tfrac{\pi}{6}D_k^3$ is the volume of a single droplet in class $k$.
 
The transport equation governing the volumetric concentration of each size class accounts for axial advection, radial turbulent diffusion, droplet breakup, and coalescence:
\begin{equation}
    v\,\frac{\partial(\tilde{n}_k\Omega_k)}{\partial z}
    + u\,\frac{\partial(\tilde{n}_k\Omega_k)}{\partial r}
    = \left(\mathcal{B}_k + \mathcal{C}_k\right)\Omega_k
    + \frac{\mathcal{D}_t}{r}\frac{\partial}{\partial r}
    \!\left(r\,\frac{\partial(\tilde{n}_k\Omega_k)}{\partial r}\right),
    \label{eq:population_balance}
\end{equation}
where $v$ and $u$ are the mean axial and radial velocity components, $\mathcal{D}_t$ is the turbulent diffusivity, which is assumed to be independent of the radial position, and $\mathcal{B}_k$ and 
$\mathcal{C}_k$ are the breakup and coalescence source terms, respectively, whose formulations are detailed in the following paragraphs.
 
\subsection*{Centreline Reduction and Similarity Analysis}

\begin{figure}
\centering
\includegraphics[scale=0.70]{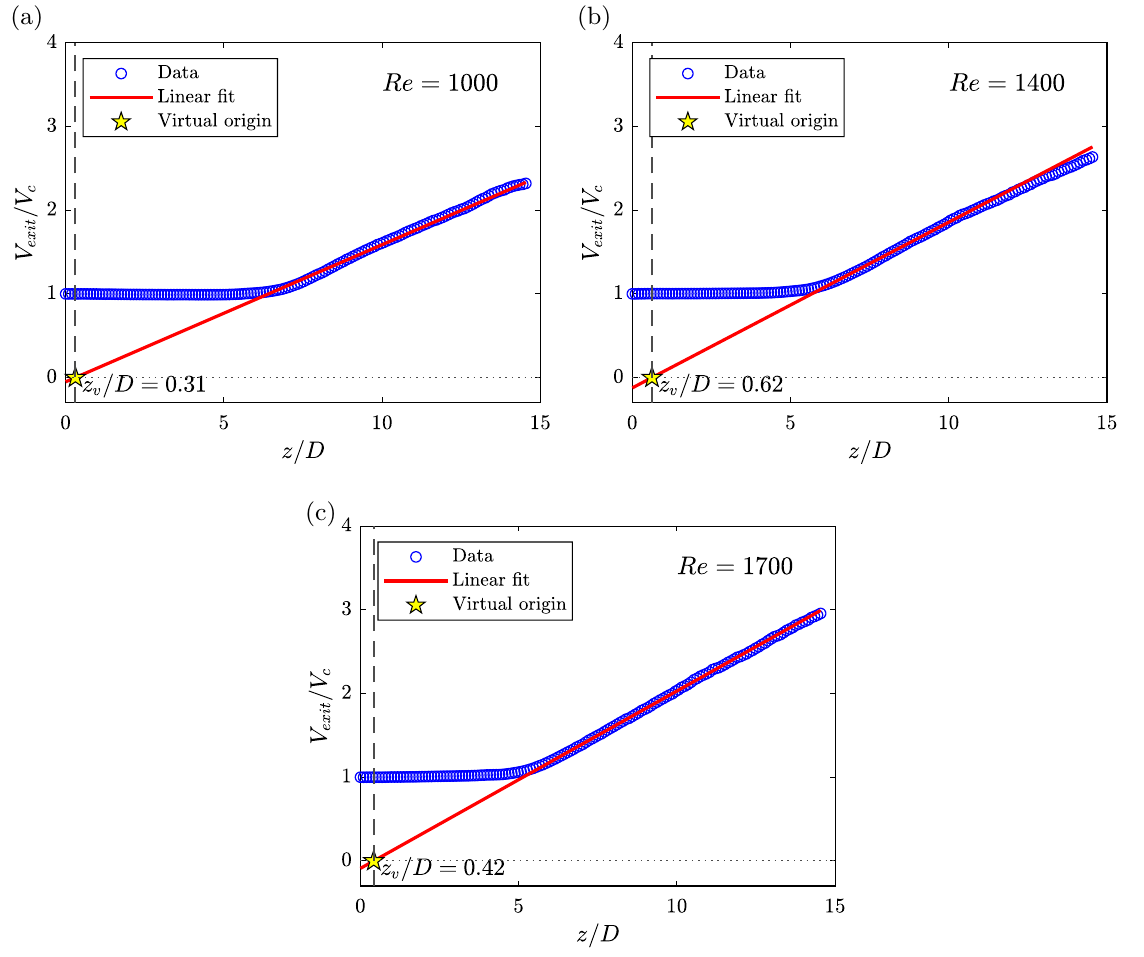}
\caption{Virtual origin for different Reynolds numbers, 
$Re = 1000$, 1400, and 1700.}
\label{fig:V_O}
\end{figure}

Experimental observations indicate that both the mean velocity and the concentration fields exhibit self-similar behaviour downstream of jet breakup for large $Re$ cases. Accordingly, the decay of the centreline velocity and the radial growth of the jet are modelled as
\begin{equation}
\frac{V_{exit}}{V_c(z)} = \frac{1}{C_m}\left(\frac{z}{D} - \frac{z_v}{D}
\right),
\qquad
r_{1/2}(z) = K (z - z_v),
\label{eq:jet_scaling}
\end{equation}
where $V_{\rm exit}$ and $D$ are the nozzle exit velocity and diameter, $z_v$ is the virtual origin, $C_m$ is the velocity decay constant, and $K$ is the jet spreading rate. Figure~\ref{fig:V_O} illustrates the estimation of the virtual origin obtained by linearly fitting the inverse centreline velocity profile ($V_{\mathrm{exit}}/V_c$). The virtual origin is determined as the point where the extrapolated linear fit intersects the abscissa. Using this procedure, the virtual origins for $Re = 1000$, 1400, and 1700 are estimated to be $z_v/D = 0.31$, 0.62, and 0.42, respectively. Since these values are all less than one nozzle diameter and are negligible compared with the overall streamwise extent of the measurements, the virtual origin is taken as $z_v = 0$ for the subsequent analysis. Accordingly, the velocity decay constant ($C_m$) and the jet spreading rate ($K$) are determined by fitting Eq.~\eqref{eq:jet_scaling} to the experimental data while assuming a virtual origin of $z_v = 0$.

\begin{figure}
\centering
\includegraphics[scale=0.65]{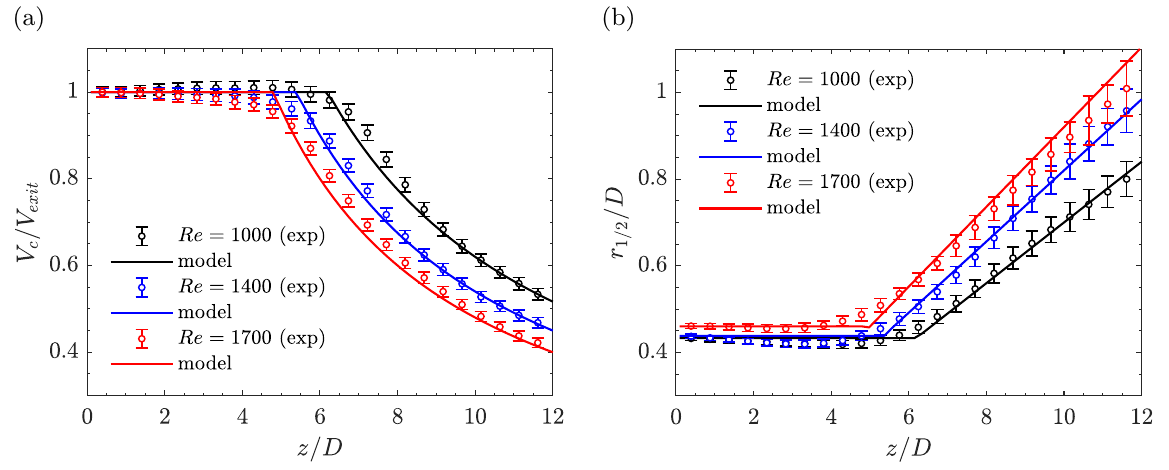}
\caption{Comparison between experimental measurements and theoretical predictions of (a) centreline velocity ($V_{c}$), normalised by the exit centreline velocity ($V_{exit}$), and (b) velocity half-width ($r_{1/2}$), normalised by the nozzle diameter ($D$), for different $\Rey$.}
\label{fig:cv_hw_theory_comp}
\end{figure}

The resulting values of constants $C_m$ and $K$, which vary with Reynolds number, are summarised in Table~\ref{table:spreading_const} and the corresponding comparisons with measurements are shown in figure~\ref{fig:cv_hw_theory_comp}.

\begin{table}
    \centering
    \caption{Experimentally obtained velocity decay constant ($C_m$) and jet spreading rate ($K$) for
    different $\Rey$.}
    \label{table:spreading_const}
    \begin{tabular}{ccc}
        \toprule
        $\Rey$ & $C_m$ & $K$ \\
        \midrule
        1000 & 6.2 & 0.070 \\
        1400 & 5.4 & 0.082 \\
        1700 & 4.8 & 0.092 \\
        \bottomrule
    \end{tabular}
\end{table}
 
Invoking similarity arguments and assuming that the normalised radial distribution of droplet concentration is unaffected by breakup or coalescence, the
two-dimensional population balance equation~\eqref{eq:population_balance} reduces to a 1D evolution equation along the jet centreline:
\begin{equation}
   \frac{d\tilde{n}_{c,k}}{dz}
    = \left[\mathcal{B}_k(z,0) + \mathcal{C}_k(z,0)\right]
      \frac{z}{V_{exit}\,C_m\,D}
    - \frac{\tilde{n}_{c,k}(z)}{z}, 
    \label{eq:centerline_ode} 
\end{equation} 
where $\tilde{n}_{c,k}$ is the centreline number density of droplets in class $k$. The first term on the right-hand side represents the combined effects of droplet breakup and coalescence, while the second term accounts for dilution arising from jet spreading and turbulent transport. The formulation of  the reduced 1D evolution equation from equation~\eqref{eq:population_balance} is given in \citet{aiyer2020coupled} and documented in Appendix~\ref{appndxA}. The combined breakup and coalescence source terms satisfy volume conservation, such that
\begin{equation}
    \sum_{k=1}^{M} \left(\mathcal{B}_k + \mathcal{C}_k\right)\Omega_k = 0.
    \label{eq:volume_conservation}
\end{equation}
 
\subsection*{Breakup Source Term}
 
The breakup source term accounts for both the formation of droplets of size $D_k$ from larger parent droplets and the loss of droplets due to their own fragmentation:
\begin{equation}
\mathcal{B}_k(z,0)
=\sum_{l\ge k}\lambda(z,0,D_l)\,\tilde{n}_{c,l}
\Big[\Phi_b(D_k,D_l)+\tilde{\Phi}_b(D_k,D_l)\Big]
-\lambda(z,0,D_k)\,\tilde{n}_{c,k},
\label{eq:breakup_source}
\end{equation}
where $\Phi_b(D_k,D_l)+\tilde{\Phi}_b(D_k,D_l)$ is the full two-daughter distribution for a breakup event of a parent droplet of class$~l$ (detailed explanation is given in Appendix~C), and the last term is the loss of class-$k$ parents due to breakup. $\Phi_b(D_k,D_l)$ is the daughter droplet probability and $\lambda$ is the breakup frequency. 
The probability of forming a daughter droplet of diameter $D_k$ from a parent droplet of diameter $D_l$, derived from surface energy considerations for binary breakup \citep{zhao2014vdrop}, is 
\begin{equation}
\Phi_b(D_k,D_l)
=
\frac{
E_{\min} + \left(E_{\max} - E(D_k)\right)
}{
\sum_{m=1}^{l-1}
\left[
E_{\min} + \left(E_{\max} - E(D_m)\right)
\right]
},
\label{eq:probability}
\end{equation}
where the surface-energy terms are
\begin{align}
E_{\max} &= \pi \sigma D_l^2 (2^{1/3} - 1), \\[4pt]
E_{\min} &= \pi \sigma \left[(D_l^3 - D_{\min}^3)^{2/3}
            + D_{\min}^2 - D_l^2 \right], \\[4pt]
E(D_k)   &= \pi \sigma \left[(D_l^3 - D_k^3)^{2/3}
            + D_k^2 - D_l^2 \right].
\end{align}
The value of $D_{\text{min}}$ is taken to be 10 $\mu$m. The breakup frequency $\lambda(D_k)$ is modelled following \citet{aiyer2019population} and is detailed in Appendix~\ref{appndxB}.
 
\subsection*{Coalescence Source Term}
 
The coalescence source term describes the net rate of change of the class-$k$ population due to binary droplet merging. It comprises a gain term, accounting for the formation of class-$k$ droplets through coalescence of all pairs $(i,j)$, and a loss term, accounting for the depletion of class-$k$ droplets through coalescence with droplets of any other class:

\begin{equation}
\mathcal{C}_k(z,0)=\frac12\sum_{i=1}^{M}\sum_{j=1}^{M}w_k\big(\tilde D_{ij}\big)\,C(D_i,D_j)\,\tilde n_{c,i}\,\tilde n_{c,j}
-\tilde n_{c,k}\sum_{j=1}^{M}C(D_k,D_j)\,\tilde n_{c,j},
\label{eq:4.11}
\end{equation}
with $\tilde D_{ij}=(D_i^3+D_j^3)^{1/3}$ and $w_k$ the volume-conserving binning weight defined in Appendix~C, where the factor of $\tfrac{1}{2}$ prevents double counting. The inclusion of both gain and loss terms, as in the breakup source term (Eq.~\ref{eq:breakup_source}), ensures a unified treatment of both mechanisms within the population balance framework.

\subsubsection*{Coalescence Rate}
 
The coalescence rate $C(D_i, D_j)$ is expressed as the product of the collision frequency and the coalescence efficiency 
\citep{abrahamson1975collision,zhao2014vdrop}:
\begin{equation}
C(D_i, D_j) = K_c\, A_{ij}
\left(v_i^2 + v_j^2\right)^{1/2} P_{ij},
\label{eq:coal_rate}
\end{equation}
where $K_c$ is an adjustable empirical coalescence coefficient analogous to $K^*$ in the breakup model, $A_{ij} = \tfrac{\pi}{4}(D_i + D_j)^2$ is the collision cross-sectional area of droplets of diameters $D_i$ and $D_j$, $v_i$ and $v_j$ are the characteristic turbulent velocities of droplets $D_i$ and $D_j$, respectively, and $P_{ij}$ is the coalescence efficiency. The droplet velocity is estimated from Kolmogorov scaling \citep{zhao2014vdrop}:
\begin{equation}
v_i = 1.03\,(\varepsilon\, D_i)^{1/3},
\label{eq:droplet_vel}
\end{equation}
where $\varepsilon$ is the local turbulent energy dissipation rate prescribed by Eq.~\ref{eq:dissipation_rate_th}.
 
\subsubsection*{Coalescence Efficiency}
 
The coalescence efficiency $P_{ij}$ represents the probability that two colliding droplets coalesce rather than rebound, and is modelled as 
\citep{prince1990bubble}:
\begin{equation}
P_{ij} = \exp\!\left(-\frac{t_{ij}}{\tau_{ij}}\right),
\label{eq:coal_eff}
\end{equation}
where $t_{ij}$ is the film drainage time and $\tau_{ij}$ is the droplet-droplet contact time. This exponential form is analogous to the breakup efficiency $P_b$, where the exponent compares a characteristic process time to a collision timescale. The contact time is given by \citep{prince1990bubble}:
\begin{equation}
\tau_{ij} = \frac{r_{ij}^{2/3}}{\varepsilon^{1/3}},
\label{eq:contact_time}
\end{equation}
where $r_{ij}$ is the equivalent radius of the colliding droplet pair, defined as \citep{prince1990bubble}:
\begin{equation}
r_{ij} = \frac{1}{2}
\left(\frac{2}{D_i} + \frac{2}{D_j}\right)^{-1}
= \frac{D_i D_j}{4(D_i + D_j)}.
\label{eq:equiv_radius}
\end{equation}
The film drainage time represents the time required for the intervening liquid film to drain from its initial thickness $h_0$ to the critical rupture thickness $h_f$
\citep{kirkpatrick1974influence}:
\begin{equation}
    t_{ij} = \left(\frac{r_{ij}^3\,\rho_c}{16\,\sigma}\right)^{1/2}
             \ln\!\left(\frac{h_0}{h_f}\right),
    \label{eq:coal_time}
\end{equation}
where $\rho_c$ is the density of the continuous phase, $\sigma$ is the interfacial tension, $h_0 = 1\times10^{-4}$~m is the initial film thickness, and $h_f = 1\times10^{-8}$~m is the critical film thickness at which rupture occurs \citep{kirkpatrick1974influence}. 
 
\subsection*{ Spatially Varying Coalescence Coefficient}
Unlike the breakup coefficient $K^*$, the coalescence coefficient $K_c$ exhibits significant spatial variation along the jet axis. Near the nozzle, the elevated oil volume fraction (holdup $\phi\approx 0.4$) and intense turbulent fluctuations promote frequent droplet-droplet collisions, enhancing coalescence. Farther downstream, both the holdup and turbulence intensity decay through entrainment and jet spreading, progressively suppressing coalescence. To capture this spatial dependence, $K_c$ is modelled as an exponentially decaying function of axial distance:
 
\begin{equation}
    K_c(z) = K_{c,\infty} + (K_{c,0} - K_{c,\infty}) \exp\!\left( -K \frac{z - z_0}{D} \right)
    \label{eq:kc_spatial}
\end{equation}
 
where $K_{c,0}$ is the near-nozzle value, $K_{c,\infty}$ is the far-field asymptote, $z_0$ is the breakup point $(z_0 = C_mD)$, and $K$ is the jet spreading rate from Eq.~\eqref{eq:jet_scaling}. The near-nozzle value $K_{c,0} = 0.2$ reflects the highly concentrated region where coalescence competes with breakup, while the far-field value $K_{c,\infty} = 5\times10^{-4}$ corresponds to the dilute far-field regime in which coalescence is subdominant, consistent with the value reported by \citet{zhao2014vdrop} for subsea oil jets. It is to be noted that variation of $K_c$ is proposed based on the present experimental observations, and its behavior is constrained to approach the asymptotic value.

\subsection{ Numerical Implementation}
 
The centreline concentration is related to the volumetric injection rate $Q_i$ as (see Appendix~\ref{appndxA})
\begin{equation}
\tilde{c}_c(z)
= \frac{Q_i\,\beta^2\,(2\,\mathrm{Sc}_T + 1)}
       {\pi\, V_c(z)\, (z-z_v)^2},
\label{eq:centerline_concentration}
\end{equation}
where $\beta^2 = (\sqrt{2}-1)/K^2$ and $\mathrm{Sc}_T = 0.7$ is the 
turbulent Schmidt number.

 The droplet number concentration is discretised into a finite number of logarithmically spaced size bins. The number of size classes, i.e. $M$, was selected after performing a grid-independence study where $M$ was varied from 20 to 60. The predicted droplet size distribution varied negligibly as $M$ increased from 40 to 60, but resulted in significantly increasing the computational cost. Thus, a value of $M=40$ was chosen as an appropriate balance between numerical accuracy and computational efficiency.
 The transport velocity used to advect the droplet population along the jet axis is the local centreline velocity $V_c(z)$, which decays as $z^{-1}$ in the self-similar region and is prescribed by Eq.~\eqref{eq:jet_scaling}. At the breakup location, corresponding to the breakup length ($z/D = C_m$), the entire droplet population is assumed to be initially contained within the size bin corresponding to the nozzle diameter $D$. This serves as the initial condition for the downstream evolution of the droplet size distribution through breakup and coalescence.
The minimum droplet diameter is specified as $D_{\min}=10\,\mu\mathrm{m}$, which is smaller than the Kolmogorov length scale for all the experimental conditions considered in the present study. This choice ensures that the smallest physically relevant droplets generated by the breakup process are adequately represented within the discretised size spectrum.
Since binary coalescence is incorporated in the population balance model, the maximum droplet diameter is taken as
$
D_{\max}=2^{1/3}D,
$
which corresponds to the largest droplet that can be formed by the coalescence of two droplets of diameter $D$. This upper limit enables the model to capture the increase in droplet size resulting from binary coalescence events. Since the solution of the population balance model requires several input parameters and constants, a summary of these quantities is provided in Table~\ref{tab:pbe_closures} for ease of reference.

\begin{table}
\centering
\caption{Summary of the closures and parameters used in the population balance model.}
\label{tab:pbe_closures}
\renewcommand{\arraystretch}{1.25}
\setlength{\tabcolsep}{6pt}
\begin{tabular}{@{}llll@{}}
\hline
\textbf{Closure} & \textbf{Parameter} & \textbf{Value} & \textbf{Source} \\
\hline
Mean velocity
    & $C_m$
    & $6.2,\;5.4,\;4.8$
    & PIV fit \\
Jet spreading
    & $K$
    & $0.070,\;0.082,\;0.092$
    & PIV fit \\
Virtual origin
    & $z_v$
    & $0$
    & Assumed \\
Dissipation
    & $C_{\varepsilon}$
    & $65$
    & Empirical \\
Breakup
    & $K^{*}$
    & $0.18$
    & Experimental fit \\
Minimum droplet size
    & $D_{\min}$
    & $10~\mu\mathrm{m}$
    & Prescribed \\
Turbulent diffusivity
    & $Sc_T$
    & $0.7$
    & \citet{Pope_2000} \\
Coalescence
    & $K_{c,0}$
    & $0.2$
    & Experimental fit \\
Coalescence
    & $K_{c,\infty}$
    & $5\times10^{-4}$
    & \citet{zhao2014vdrop} \\
Initial film thickness
    & $h_0$
    & $1\times10^{-4}$~m
    & \citet{kirkpatrick1974influence} \\
Critical film thickness
    & $h_f$
    & $1\times10^{-8}$~m
    & \citet{kirkpatrick1974influence} \\
Size bins
    & $M$
    & $40$
    & Numerical \\
Initial droplet size
    & $D_M$
    & $D$
    & Experimental \\
\end{tabular}
\end{table}


The system of coupled ordinary differential equations (Eqs.~\ref{eq:centerline_ode}, \ref{eq:centerline_concentration}) is numerically solved along with  Eq.~\ref{eq:volume_conservation} starting from $z = C_m D$ (breakup length), with all injected volume initially assigned to the droplet class, $D_M \approx D$.
The coupled system of ordinary differential equations is integrated by marching in the axial direction using a first-order explicit Euler scheme. The spatial domain, from $z = C_m D$ to $z = 0.3$~m, is divided into $N_z = 5000$ uniformly spaced points. At each axial location, the local dissipation rate $\varepsilon(z)$ and coalescence coefficient $K_c(z)$ are evaluated, and the number density in each bin is updated sequentially by breakup (using the breakup frequency $\lambda(D_k)$ and daughter-size probability density) and coalescence ($C(D_i,D_j)$ with the local $K_c(z)$). In both breakup and coalescence, daughter or merged droplet volumes falling between two bin sizes are redistributed onto the two neighbouring bins by linear interpolation in volume space, conserving oil volume exactly. The numerical procedure adopted to enforce volume conservation during the integration of the population balance equations is described in detail in Appendix~C. The scheme was verified to be numerically stable and grid-independent by confirming that further refinement of the axial step size or the number of bins did not appreciably change the resulting droplet size distribution or the total oil volume, which is conserved to within numerical round-off at every axial location.
Note that the Breakup and coalescence source terms are updated at each axial location, which depends on the local turbulent dissipation rate. The axial variation of the turbulent dissipation rate is prescribed as
\begin{equation}
\varepsilon(z) =
C_\varepsilon \frac{V_{exit}^3}{D}
\left(\frac{z}{D} - \frac{z_v}{D}\right)^{-4},
\label{eq:dissipation_rate_th}
\end{equation}
where $C_\varepsilon = 65$ is an empirical constant \citep{aiyer2020coupled}.
 
The turbulent dissipation rate is calculated using the velocity field obtained from the PIV data following the energy spectra method of \citet{xu2013accurate}. In this approach, the one-dimensional longitudinal energy spectrum $\tilde{E}_{11}(k_1)$ is first computed from the PIV velocity data using a fast Fourier transform after subtracting the ensemble mean velocity and applying a windowing function to the fluctuating velocity field. Because the PIV interrogation procedure is mathematically equivalent to a spatial filtering operation \citep{xu2013accurate}, the measured spectrum $\tilde{E}_{11}(k_1)$ is related to the true longitudinal energy spectrum $E_{11}(k_1)$ by
\begin{equation}
\tilde{E}_{11}(k_1) = |\hat{G}(k_1)|^2\, E_{11}(k_1),
\label{eq:spectra_filter}
\end{equation}
where $\hat{G}(k_1)$ is the Fourier transform of the interrogation window filter function $G$, analogous to the filtering operation in large eddy simulations \citep{Pope_2000} . Inverting Eq.~(\ref{eq:spectra_filter}) corrects for the averaging effect introduced by the finite interrogation window size $\Delta$, yielding a corrected energy spectrum
\begin{equation}
E_{11}(k_1) = \frac{\tilde{E}_{11}(k_1)}{|\hat{G}(k_1)|^2}.
\label{eq:spectra_corrected}
\end{equation}
In the inertial subrange, the corrected spectrum follows Kolmogorov's energy cascade scaling
\begin{equation}
E_{11}(k_1) = \frac{18}{55}\, C_k\, \varepsilon^{2/3}\, k_1^{-5/3},
\label{eq:spectra_inertial}
\end{equation}
where $C_k = 3/2$ is the Kolmogorov constant. Fitting the $-5/3$ slope to the inertial subrange of the corrected spectrum then yields the turbulent dissipation rate as
\begin{equation}
\varepsilon = \left(\frac{55}{18\,C_k}\, k_1^{5/3}\, E_{11}(k_1)\right)^{3/2}.
\label{eq:epsilon_spectra}
\end{equation}
The spectral correction in Eq.~(\ref{eq:spectra_filter}) depends only on the interrogation-window transfer function and is therefore independent of $C_k$. The dissipation rate is subsequently estimated by fitting the corrected spectrum to the Kolmogorov $-5/3$ scaling, which introduces a dependence on the Kolmogorov constant $C_k$. While variations in $C_k$ may marginally affect the absolute value of $\varepsilon$, a consistent value of $C_k$ is used throughout the present analysis, ensuring that the relative comparison between different Reynolds numbers remains consistent.

\begin{figure}
\centering
\includegraphics[scale=0.75]{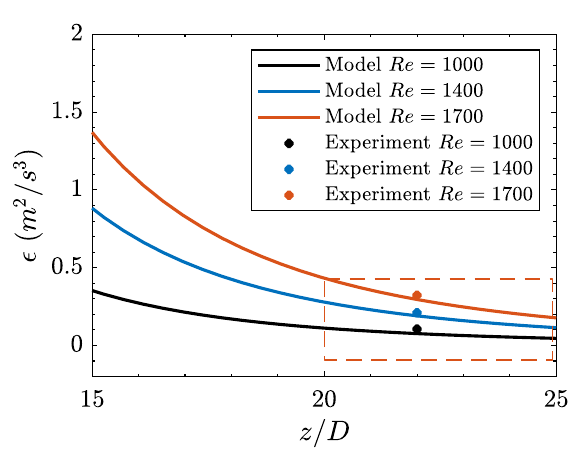}
\caption{Turbulent dissipation rate at different Reynolds numbers, 
$Re = 1000$, 1400, and 1700.  The rectangle indicates the axial extent of the window over which the dissipation rate is estimated from the PIV energy spectra.}
\label{fig:dissipation}
\end{figure}

The modified spectral curve-fit method of \citet{xu2013accurate} is less sensitive to the PIV interrogation window size $\Delta$ than direct velocity-gradient methods, which significantly underestimate $\varepsilon$ when $\Delta$ exceeds the Kolmogorov length scale $\eta_k = (\nu^3/\varepsilon)^{1/4}$, since the spatially filtered velocity gradients cannot resolve turbulent motions at scales smaller than $\Delta$. The spectral method is, however, applicable only in the far field ($z/D \gtrsim 15$), where the turbulence is sufficiently isotropic for Kolmogorov's similarity hypothesis to hold, and a well-defined inertial subrange is present in the energy spectrum. In the near-nozzle region, strong mean shear and anisotropic turbulent structures violate the local isotropy assumption, rendering the spectral estimate unreliable. The PIV-based dissipation rates for $\Rey = 1000$, 1400, and 1700 are therefore compared with the 
theoretical prediction of Eq.~\eqref{eq:dissipation_rate_th} exclusively in the far field. As shown in figure~\ref{fig:dissipation}, good agreement is obtained across all Reynolds numbers considered, validating the use of the theoretical dissipation rate~\eqref{eq:dissipation_rate_th} throughout the 1D ODE model.
 
\subsection{ Model Predictions and Comparison with Experiments}
\label{sec:model_results}
The model predicts the evolution of droplet size distributions along the jet centreline by accounting for turbulent transport, breakup, and coalescence processes, enabling direct comparison with experimentally measured droplet statistics. In the experiments, the droplet size distribution is obtained from an interrogation window of dimensions $5D \times 10D$ (height $\times$ width). Owing to the finite number of droplets required for reliable statistical convergence, it is not practical to determine the droplet size distribution at a single axial location. Moreover, if a smaller window were used, droplets with sizes comparable to the window dimensions would intersect the window edges and would not be sampled reliably. Instead, the droplet statistics are evaluated over the entire interrogation window and are associated with its geometric centre for comparison with the model predictions.
In contrast, the 1D ODE model predicts the droplet number density at discrete axial locations. Therefore, the model predictions are compared with the experimentally measured distribution at the centre of the interrogation window, which represents the average droplet population within that measurement region.
 
\begin{figure}
\centering
\includegraphics[scale=0.60]{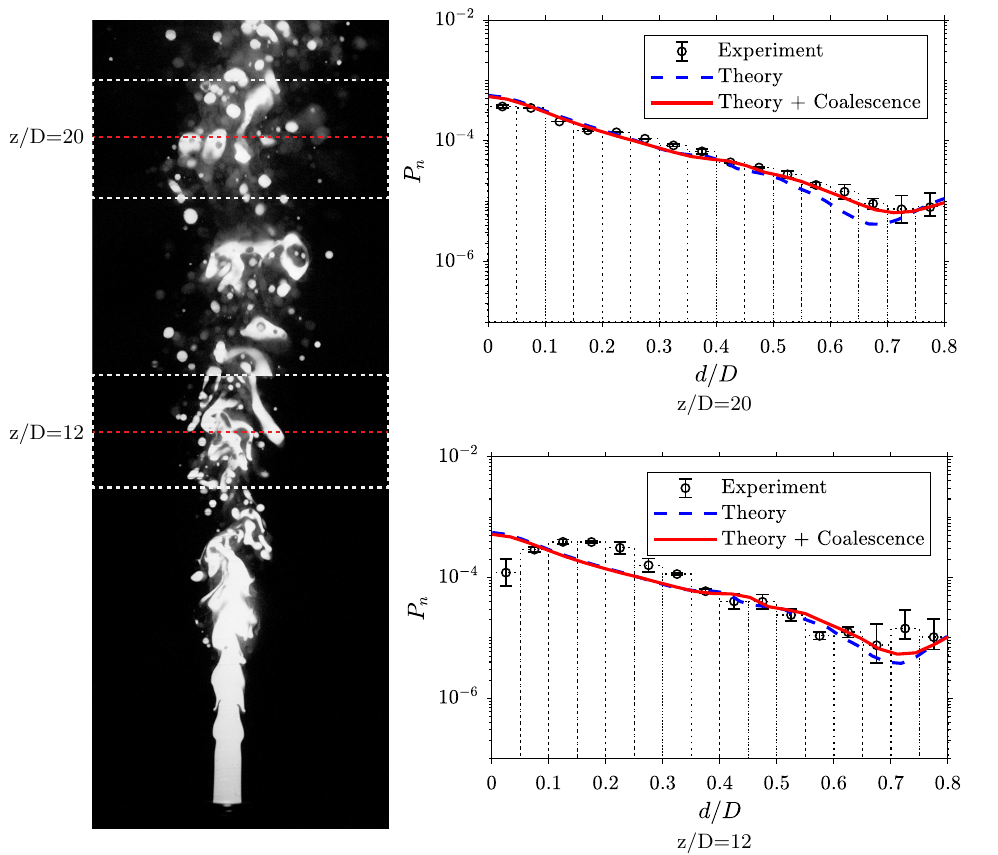}
\caption{Droplet size distribution for $\Rey = 1000$ at axial positions $z/D = 12$ and $20$. Left panel: Instantaneous PLIF image with the measurement planes indicated by red dashed lines. The white dashed lines denote the axial extent of the measurement window over which the size distribution is evaluated. Right panel: number probability density $P_n$ at two axial locations. The blue dashed curve represents the theoretical prediction accounting for breakup only, while the red curve includes both breakup and coalescence. Error bars indicate the standard deviation from three repetitions.}
\label{fig:size_dist1000}
\end{figure}
 
\begin{figure}
\centering
\includegraphics[scale=0.60]{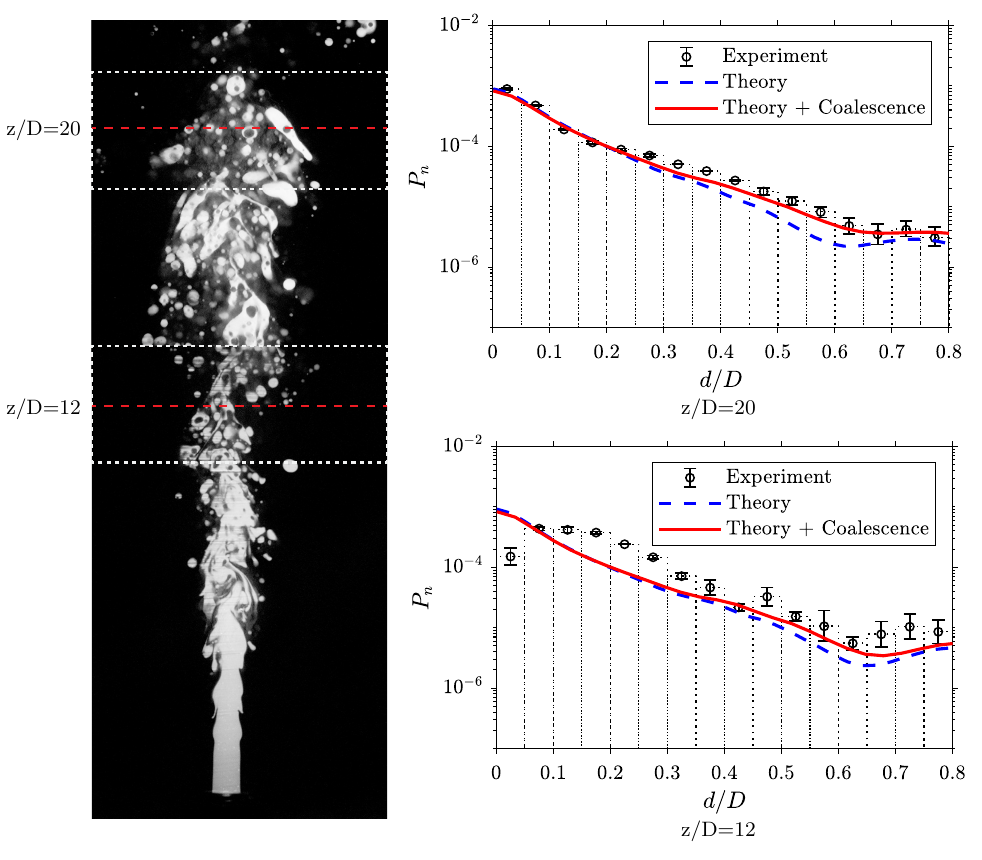}
\caption{Droplet size distribution for $\Rey = 1400$ at axial positions $z/D = 12$ and $20$. Left panel: Instantaneous PLIF image with the measurement planes indicated by red dashed lines. The white dashed lines denote the axial extent of the measurement window over which the size distribution is evaluated. Right panel: number probability density $P_n$ at two axial locations. The blue dashed curve represents the theoretical prediction accounting for breakup only, while the red curve includes both breakup and coalescence. Error bars indicate the standard deviation from three repetitions.}
\label{fig:size_dist1400}
\end{figure}
 
\begin{figure}
\centering
\includegraphics[scale=0.60]{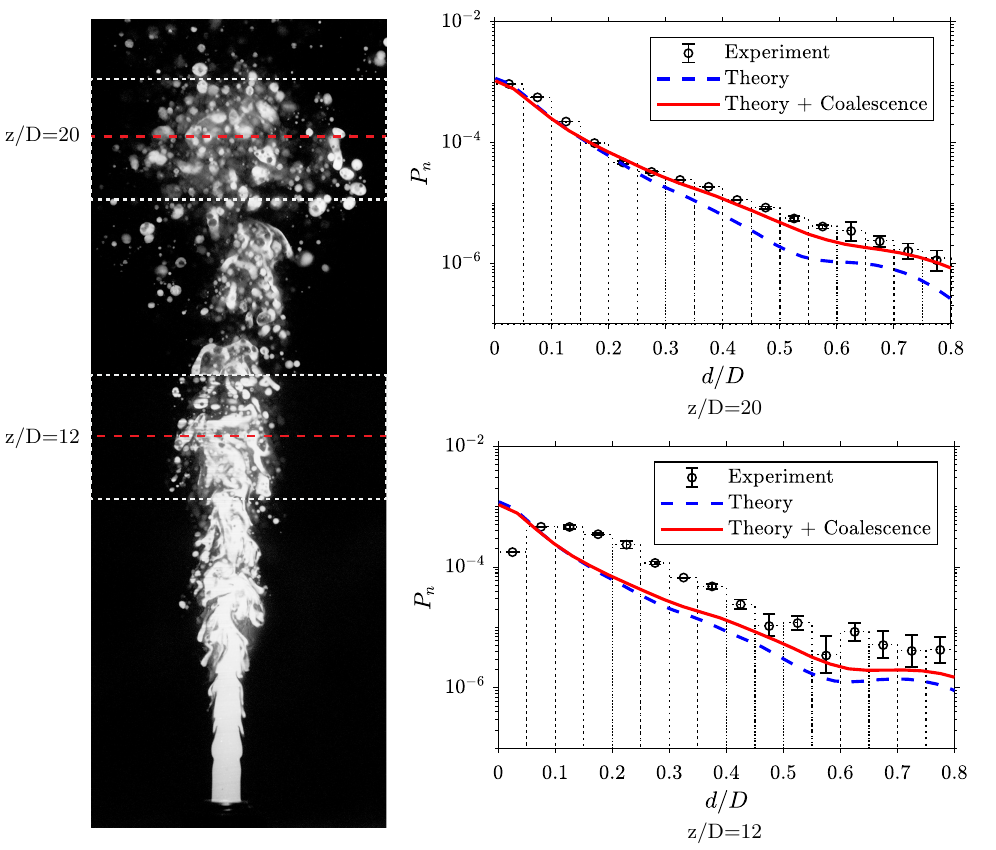}
\caption{Droplet size distribution for $\Rey = 1700$ at axial positions $z/D = 12$ and $20$. Left panel: Instantaneous PLIF image with the measurement planes indicated by red dashed lines. The white dashed lines denote the axial extent of the measurement window over which the size distribution is evaluated. Right panel: number probability density $P_n$ at two axial locations. The blue dashed curve represents the theoretical prediction accounting for breakup only, while the red curve includes both breakup and coalescence. Error bars indicate the standard deviation from three repetitions.}
\label{fig:size_dist1700}
\end{figure}
 
Figures~\ref{fig:size_dist1000}, \ref{fig:size_dist1400}, and 
\ref{fig:size_dist1700} present a comparison between the number probability density, $P_n(d)$, obtained from the PLIF measurements and the predictions of the 1D ODE model at two axial locations, $z/D = 12$ and $z/D = 20$, for $\text{Re} = 1000$, 1400, and 1700. These Reynolds numbers correspond to  turbulent jet conditions. The figures also compare the original theoretical model, which accounts only for droplet breakup, with the present extended formulation that incorporates both breakup and coalescence mechanisms.
 
Overall, the model predictions capture the experimentally measured droplet size distributions reasonably well, particularly in the downstream region. However, a noticeable deviation between the model and experiments is observed in the initial stages of the jet development. This discrepancy arises because the model assumes that the liquid jet has already undergone complete disintegration into droplets, whereas in the experiments the jet core continues to fragment in the near field. In addition, the breakup and coalescence closures assume locally homogeneous and isotropic turbulence and eddy-collision-driven breakup, which are only approximate in the developing near field, where the largest droplets exceed the integral length scale (see the model limitations in Section~\ref{sec:limitations}). The jet is observed to be fully disintegrated only farther downstream, around $z/D \approx 20$, where the model assumptions become more consistent with the physical flow conditions. A noticeably improved agreement with the experiments is obtained when coalescence is included in the model. The root-mean-square error (RMSE) between the experimental and model-predicted $P_n(d)$ distributions at $z/D=20$ is found to be 7$\%$, 9$\%$, and 9.5$\%$ for $Re$=1000, 1400, and 1700, respectively. The inclusion of coalescence introduces an additional pathway for the formation of larger droplets: when two droplets of diameters $D_i$ and $D_j$ collide and merge, they form a droplet of diameter $D_k = (D_i^3 + D_j^3)^{1/3}$, thereby repopulating the larger size classes of the distribution. As a result, the model with coalescence predicts a broader distribution and a slightly higher probability of intermediate and larger droplets, consistent with the experimental observations. The influence of coalescence is found to be more pronounced in the far field ($z/D = 20$) compared with the near field ($z/D = 12$). This behaviour can be attributed to the spatial variation of turbulent kinetic energy along the jet. In the near field, the turbulent dissipation rate is high and promotes frequent breakup events, which suppress the relative influence of coalescence. Although the droplet concentration is higher in this region and the absolute coalescence rate can therefore be significant, the ratio of coalescence to breakup remains relatively small. As the jet evolves downstream, the turbulent kinetic energy decreases, reducing the breakup rate and increasing the relative contribution of coalescence, which enhances the formation of larger droplets in the far field.
 
Another notable feature is the presence of a weak bimodal character in the droplet size distribution at $z/D = 12$, which is observed in both the experimental data and the model predictions. This bimodal behaviour gradually diminishes downstream, and the distribution transitions to a monotonically decreasing single mode at $z/D = 20$. The origin of this dual-mode behaviour can be linked to the breakup mechanism employed in the model, which is based on surface energy considerations. Within this framework, the formation of equal-sized daughter droplets from a parent droplet is statistically less probable, and breakup events tend to produce droplets of unequal sizes. At these Reynolds numbers, the turbulent dissipation, and therefore the available surface energy for fragmentation, is lower than in extremely high $\Rey$ jets. Consequently, breakup events produce a mixture of smaller and intermediate-sized droplets, giving rise to the observed dual-mode distribution. Further downstream, continuous breakup and coalescence interactions redistribute the droplet population, smoothing the intermediate peak and ultimately leading to a single-mode distribution.
\begin{figure}
\centering
\includegraphics[scale=0.70]{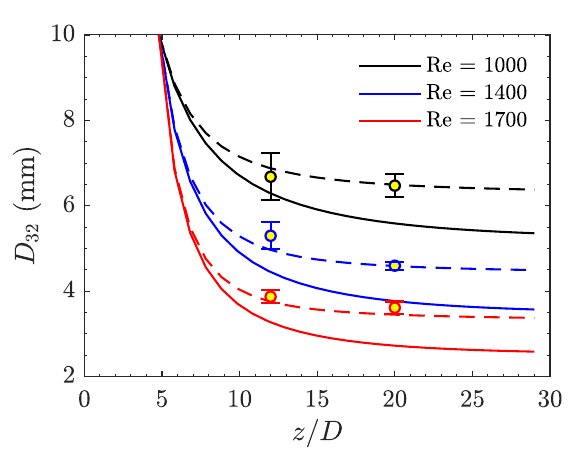}
\caption{Variation of the predicted Sauter mean diameter, $D_{32}$, with axial distance for different Reynolds numbers. Solid lines denote model predictions considering breakup only, while dashed lines represent the extended model that includes the effects of droplet coalescence. Circles with error bars denote experimental values of $D_{32}$ at $z/D = 12$ and 20.}
\label{fig:d32_variation}
\end{figure}
 
The axial evolution of the Sauter mean diameter, $D_{32}$, predicted theoretically and measured experimentally for the three Reynolds numbers is presented in figure~\ref{fig:d32_variation}. The model predictions accounting for breakup only and for both breakup and coalescence are compared against the experimental data. In all cases, $D_{32}$ decreases monotonically with increasing axial distance. This trend reflects the progressive fragmentation of droplets as the jet evolves downstream, where turbulent shear and interfacial instabilities continuously break larger droplets into smaller ones. Since $D_{32}$ is defined as the ratio of the third to the second moment of the droplet size distribution,
\begin{equation}
D_{32} = \frac{\sum_i n_i D_i^3}{\sum_i n_i D_i^2},
\end{equation}
it represents a surface-area-weighted mean diameter and therefore provides a physically meaningful measure of the characteristic droplet size governing interfacial transport processes.
 
A comparison between the two model predictions shows that the inclusion of coalescence results in systematically larger values of $D_{32}$ relative to the breakup-only model. This behaviour arises because coalescence acts as a counteracting mechanism to breakup by merging smaller droplets to form larger ones, thereby repopulating the higher diameter classes of the distribution. Consequently, the net reduction in droplet size is moderated when coalescence is considered. In addition, the model including coalescence reaches its asymptotic $D_{32}$ value more rapidly along the axial direction. Physically, this indicates that the competing effects of breakup and coalescence quickly establish a dynamic balance in the droplet population, leading to a quasi-equilibrium droplet size distribution further downstream. The experimentally obtained $D_{32}$ values at two axial locations, $z/D = 12$ and $z/D = 20$, show good agreement with the theoretical model incorporating coalescence.  However, minor discrepancies at $z/D = 12$ are attributed to incomplete jet disintegration in this near-field region and to  closure limitations discussed in Section~\ref{sec:limitations}.

The ratio of coalescence to breakup, quantified as $R(z) = \sum_k |\mathcal{C}_k(z,0)|/\sum_k |\mathcal{B}_k(z,0)|$, remains relatively small, ranging from $0.08$ to $0.2$ for all cases considered. Thus, breakup remains the dominant mechanism, with coalescence occurring at approximately one order of magnitude lower rate. This is comparable to the observations of \citet{kewalramani2022experimental}, who reported that fragmentation events outnumbered collision events by approximately one order of magnitude in immiscible liquid--liquid jets. Despite its smaller magnitude, inclusion of coalescence improves the prediction of the droplet size distributions and Sauter mean diameter, particularly for the intermediate to larger droplet sizes.

\subsection{Limitations of the closure}
\label{sec:limitations}
The breakup and coalescence source terms presented in the above sections rely on a local, homogeneous-isotropic-turbulence closure: the breakup frequency $\lambda(D_k)$ (Appendix~B) and the coalescence rate $C(D_i,D_j)$ (Eq.~4.12) are both built from the characteristic eddy timescale $\tau_{b,k}=\varepsilon^{-1/3}D_k^{2/3}$, which is appropriate only when the droplet diameter $D_k$ is small compared with the local integral length scale $L(z)$ of the turbulence. When $D_k\gtrsim L(z)$, the relevant timescale is instead $\tau=L/u'=\varepsilon^{-1/3}L^{2/3}$, which is generally different from $\tau_{b,k}$.

Since the energy-containing eddies in a round jet are bounded by the local jet width, the integral scale scales with the jet half-width, $L\propto r_{1/2}$, with a prefactor smaller than unity \citep{Pope_2000,wygnanski1969some}; we therefore use $r_{1/2}(z)$ as an upper-bound estimate of $L(z)$. Using the velocity half-width $r_{1/2}(z)$ (Figs.~8b, 13b) as a proxy for $L(z)$, near the breakup location ($z/D\approx5$--6), $L\approx r_{1/2}\approx4$-5\,mm. The largest droplet class permitted by the model, $D_{max}=(2)^{1/3}D\approx12.6$\,mm, is therefore larger than $L$ in this region, so for these classes $\tau$ rather than $\tau_{b,k}$ is the physically appropriate breakup timescale. Farther downstream, where $r_{1/2}> D$ ($z/D\gtrsim20$), $L$ exceeds the droplet sizes of interest and the local isotropic closure is consistent. The larger discrepancy between the model predictions and measurements at $z/D=12$ (Section~4.2) may be attributed not only to incomplete jet disintegration but also to the reduced validity of the closure for the largest droplet classes in this near-field region. These classes constitute only a small fraction of the droplet population, as reflected by the larger error bars at the corresponding diameters in Figs.~20--22.

A related limitation concerns the residence time of a droplet relative to the breakup/coalescence timescale itself. The source terms $\mathcal{B}_k$ and $\mathcal{C}_k$ are evaluated locally, using the instantaneous $V_c(z)$ and $\varepsilon(z)$ at each axial station, which implicitly assumes that a droplet undergoing breakup or coalescence experiences an approximately constant $V_c$ and $\varepsilon$ over the duration of the event. This assumption is most restrictive for coalescence of large droplets, where the controlling timescale is the film-drainage time. For a pair of large droplets, $D_i=D_j=6$\,mm,  the film-drainage time from Eq.~(4.17) is $t_{ij}\approx8.5$\,ms. For $Re=1700$ ($V_{exit}=2.2$\,m\,s$^{-1}$, $C_m=4.8$) at $z/D=20$, the centreline velocity is $V_c\approx0.53$\,m\,s$^{-1}$, so the pair travels $\Delta z=V_c\,t_{ij}\approx4.5$\,mm $\approx0.45D$ during one coalescence event. Over this distance, $\varepsilon\propto z^{-4}$ decreases by about 9\% (from $z/D=20$ to $20.45$). Hence, $V_c$ and $\varepsilon$ are not strictly constant during the coalescence of the large droplets, and the locally evaluated source terms are only approximate. The effect is much weaker for smaller and intermediate-sized droplets. For a pair with $D_i=D_j\approx3.78$\,mm (Sauter mean diameter), $t_{ij}\approx4.2$\,ms and $\Delta z=V_c\,t_{ij}\approx2.2$\,mm $\approx0.22D$, over which $\varepsilon$ decreases by only about 4\% (from $z/D=20$ to $20.22$). Thus, this approximation becomes more accurate for smaller droplets.

The above limitations mainly affect the largest droplet classes
near and just downstream of the breakup region. The overall agreement
between the model and the measurements (RMSE of 7--9.5\,\% at
$z/D=20$, Section~4.2) indicates that the present model provides a
reasonable approximation for most of the droplet population. Replacing $D_k$ by $\min(D_k,L)$ in the eddy velocity and time scale accounts for the finite integral length scale $L$, but it does not capture the breakup physics of these droplets, which, for $D_k>L$, are deformed mainly by large-scale strain and mean shear rather than by eddy collisions. A more accurate prediction of the large-droplet tail therefore requires a shear-based breakup kernel for these classes, together with a history-dependent source term that accounts for the finite residence time during a breakup or coalescence event. These developments are beyond the scope of the present study and are
identified as directions for future work.

\subsection{ Effect of Continuous Phase Viscosity on Droplet Size Distribution}
\label{sec:conti_phase_visc}
\begin{figure}
\centering
\includegraphics[scale=0.70]{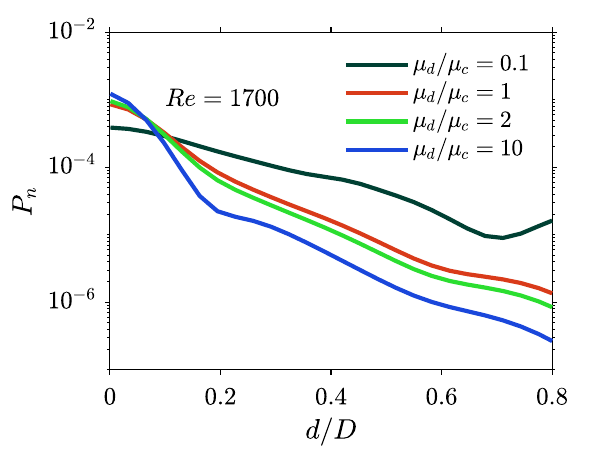}
\caption{Theoretical prediction of the number probability density for different viscosity ratios ($\mu_d/\mu_c$) at $\text{Re} = 1700$. }
\label{fig:cont_phase_viscosity}
\end{figure}
Figure~\ref{fig:cont_phase_viscosity} illustrates the effect of continuous phase viscosity on the droplet size distribution (DSD), obtained by systematically varying $\mu_d/\mu_c$ while keeping all other parameters fixed at $\Rey = 1700$ and $\We = 1229$. The results are presented in terms of the viscosity ratio $\mu_d/\mu_c$, which provides a convenient measure of the relative importance of dispersed-to-continuous-phase viscosity.
 
 The continuous phase viscosity has a pronounced influence on the shape of the DSD (figure~\ref{fig:cont_phase_viscosity}). At low values of viscosity ratio, corresponding to a relatively high continuous phase viscosity, the distribution becomes progressively flatter, indicating a greater proportion of larger droplets surviving in the downstream region of the jet. In contrast, at comparatively high values of $\mu_d/\mu_c$, corresponding to a lower continuous phase viscosity, the distribution shifts markedly toward smaller droplet sizes, signifying enhanced breakup activity.
 
This behaviour can be attributed to the stabilising role of the continuous phase viscosity on the jet dynamics. A higher continuous phase viscosity damps the growth of Kelvin-Helmholtz (KH) instabilities at the jet interface, suppressing the formation of coherent roll-up structures that would otherwise promote droplet breakup. Furthermore, the elevated viscosity dissipates turbulent kinetic energy more rapidly, leading to a lower level of sustained turbulent fluctuations in the flow. Since droplet breakup in turbulent jets is primarily driven by the local turbulent dissipation rate, the reduced fluctuation energy directly weakens the breakup mechanism, allowing larger droplets to persist further downstream.


\section{Conclusion} 
\label{sec:conclusion}
This study presents a combined experimental and theoretical investigation of immiscible liquid-liquid jet dynamics and droplet size evolution across the transitional to  turbulent regimes. Simultaneous PIV and PLIF measurements, enabled by a refractive index-matched silicone oil-sugar water fluid pair, provided time-resolved, high-quality velocity and concentration fields that formed the quantitative basis of the analysis.
 
The mean flow characterisation revealed that the jet breakup length exhibits a non-monotonic dependence on Reynolds number, increasing from the transitional to the lower atomisation regime before decreasing monotonically at higher Reynolds numbers, while the velocity half-width and post-breakup spreading rate increase due to intensified turbulent mixing. The centreline velocity decay follows the classical $z^{-1}$ scaling and the velocity half-width grows linearly with axial distance downstream of the breakup point, confirming self-similar behaviour in the  turbulent cases. Proper orthogonal decomposition (POD) analysis of the PLIF snapshots identified a clear and systematic transition in the dominant instability mechanism with increasing Reynolds number: axisymmetric Rayleigh-Plateau modes govern breakup at $Re = 125$, sinuous and varicose modes coexist in competition at $Re = 600$, and Kelvin-Helmholtz-driven shear-layer roll-up dominates at $Re = 1000$ and beyond.  This modal transition is directly reflected in the measured droplet size distributions.
 
The droplet size distributions exhibit a strong dependence on Reynolds number. In the transitional regime ($Re = 600$), the volume probability density $P_v(d)$ is distinctly bimodal, with a secondary peak at large diameters attributed to droplets produced by Rayleigh-Plateau instability of the liquid column that survive under the relatively weak turbulent stresses of this regime. As the Reynolds number increases and the jet transitions to the atomisation regime, turbulent dissipation intensifies and shear-induced breakup becomes dominant, collapsing the distribution to a single mode with progressively smaller characteristic droplet sizes. A weak bimodal character is also observed in the number probability density $P_n(d)$ near the nozzle at  turbulent conditions, which is reproduced by the population balance model and attributed to the surface-energy-based breakup mechanism that statistically favours
the production of unequal-sized daughter droplets at moderate dissipation levels. This dual-mode signature gradually diminishes downstream as breakup and coalescence interactions redistribute the droplet population toward a single-mode equilibrium distribution.
 
The one-dimensional population balance model, extended beyond the formulation of \citet{aiyer2020coupled} to include droplet coalescence, provides good agreement with the experimentally measured number probability densities for the  turbulent cases ($Re = 1000$, $1400$, and $1700$). The inclusion of coalescence systematically improves predictions at the large-droplet end of the distribution, where the breakup-only model under-predicts droplet concentrations. Coalescence effects are found to be more pronounced in the far field ($z/D = 20$) compared with the near field ($z/D = 12$), where the turbulent dissipation rate has decayed sufficiently to reduce the relative frequency of breakup events and allow merging collisions to become significant. The model also confirms that the Sauter mean diameter $D_{32}$ decreases monotonically with axial distance, with the coalescence-inclusive model predicting consistently larger $D_{32}$ values and a more rapid approach to asymptotic equilibrium compared with the breakup-only formulation, reflecting the dynamic balance established between the two competing mechanisms.
 
Parametric analysis of the continuous phase viscosity further demonstrates that increasing $\mu_c$ stabilises the jet by damping Kelvin-Helmholtz instabilities at the jet interface and accelerating the dissipation of turbulent kinetic energy. This reduces the level of sustained turbulent fluctuations and, consequently, the local breakup frequency, resulting in a flatter droplet size distribution with a greater proportion of larger droplets persisting downstream.
 
Taken together, these results establish the dominant roles of turbulent dissipation, interfacial instability modes, and droplet coalescence in controlling droplet size evolution in immiscible liquid-liquid jets, and provide a validated experimental data set and extended modelling framework applicable to environmental and industrial multiphase jet systems such as subsea oil spills, spray atomisation, and liquid-liquid extraction processes.

\section*{Acknowledgements:} 
 LMS gratefully acknowledges the Ministry of Education (MoE), India, for the Prime Minister's Research Fellowship (PMRF). LDC wishes to thank IIT Hyderabad for its support in the form of a seed grant. 
The authors acknowledge the financial support provided by the Fund for Improvement of Science and Technology Infrastructure (FIST) of India (Grant number: SR/FST/ETI-397/2015/(C)) and the JICA Friendship Project. The authors thank the anonymous reviewers for their constructive comments and suggestions to improve this manuscript.
\section*{Data Availability Statement}
The data that support the findings of this study are available from the corresponding author upon reasonable request.
\appendix
\section{ Derivation of the One-Dimensional Population Balance Equation}
\label{appndxA}

The detailed derivation of the one-dimensional population balance equation is provided in \citet{aiyer2020coupled}. For completeness, we briefly present the formulation here. The droplet population is discretized into $M$ bins according to droplet diameter $D_k$, $k = 1,\dots,M$, and the total mean droplet concentration is related to the concentration in each bin through
 
\begin{equation}
\tilde{c}(z,r) = \sum_{k=1}^{M} \Omega_k\, \tilde{n}_k(z,r),
\label{eq:total_conc}
\end{equation}
 
where $\tilde{n}_k$ is the (RANS-averaged) number density of droplets of diameter $D_k$ per unit volume of fluid and $\Omega_k = (\pi/6) D_k^3$ is the volume of a single droplet in bin $k$. The transport equation for the mean concentration of bin $k$, including advection, breakup, and radial turbulent diffusion, is written as
 
\begin{equation}
 v\,\frac{\partial(\tilde{n}_k\Omega_k)}{\partial z}
    + u\,\frac{\partial(\tilde{n}_k\Omega_k)}{\partial r}
    = \left(\mathcal{B}_k + \mathcal{C}_k\right)\Omega_k
    + \frac{\mathcal{D}_t}{r}\frac{\partial}{\partial r}
    \!\left(r\,\frac{\partial(\tilde{n}_k\Omega_k)}{\partial r}\right),
\label{eq:bin_transport}
\end{equation}
 
where $v$ and $u$ are the mean axial and radial velocities, respectively, $(\mathcal{B}_k + \mathcal{C}_k)\Omega_k$ is the source term describing the net rate of change of the bin-$k$ concentration due to droplet breakup and coalescence, and the eddy diffusivity $\mathcal{D}_t$ is taken to depend on $z$ only (i.e., independent of radial position). The individual droplet rise velocity is neglected relative to the centerline jet velocity in the near-nozzle region, and molecular diffusion is neglected since $\mathcal{D}_t$ is much larger than the molecular diffusivity at the Reynolds numbers considered.
  
The mean velocity is obtained from a classical eddy-viscosity closure. Conservation of mass and momentum for a round turbulent jet, written in polar coordinates under the boundary-layer approximation, read
 
\begin{equation}
\frac{\partial v}{\partial z} + \frac{1}{r}\frac{\partial(ru)}{\partial r} = 0,
\label{eq:continuity}
\end{equation}
\begin{equation}
v\,\frac{\partial v}{\partial z} + u\,\frac{\partial v}{\partial r}
= \nu_T\,\frac{1}{r}\frac{\partial}{\partial r}\left(r\,\frac{\partial v}{\partial r}\right),
\label{eq:momentum}
\end{equation}
 
where $\nu_T$ is the (z-dependent) eddy viscosity, and viscous diffusion by the molecular viscosity has again been dropped at the high Reynolds numbers of interest.
 
Seeking a self-similar mean axial velocity profile,
 
\begin{equation}
v(z,r) = V_c(z)\, f(\eta), \qquad \eta = \frac{r}{z - z_v},
\label{eq:velocity_similarity}
\end{equation}
 
with $V_c(z)$ the centerline velocity and $z_v$ the virtual origin of the jet, the radial velocity consistent with continuity, Eq.~\eqref{eq:continuity}, follows by integrating Eq.~\eqref{eq:continuity} radially and substituting the similarity form \eqref{eq:velocity_similarity}. Inserting both into the momentum equation, Eq.~\eqref{eq:momentum}, and assuming $\nu_T$ independent of $r$, one obtains an ordinary differential equation for $f(\eta)$ whose bounded, monotonically decreasing solution is
 
\begin{equation}
f(\eta) = \frac{1}{\left(1 + \beta^2 \eta^2\right)^2},
\label{eq:velocity_profile}
\end{equation}
 
with $\beta$ a dimensionless constant related to the jet spreading rate $K$ through $\beta^2 = (\sqrt{2} - 1)/K^2$. From the self-similar solution, the downstream evolution of the centerline velocity, the jet half-width, and the mean turbulent dissipation rate($\varepsilon$) can be deduced as
 
\begin{equation}
\frac{V_{exit}}{V_c(z)} = \frac{1}{C_m}\left(\frac{z}{D} - \frac{z_v}{D}
\right),
\qquad
r_{1/2} = K\left(z - z_v\right),
\qquad
\frac{\varepsilon D}{V_{exit}^3} = C_{\varepsilon}\left(\frac{z}{D} - \frac{z_v}{D}\right)^{-4},
\label{eq:centerline_laws}
\end{equation}
 
where $C_m$, $K$, and $C_{\varepsilon}$ are empirically determined constants, $D$ is the nozzle diameter, and $z_v$ is the virtual origin.
  
We next seek a similarity description of Eq.~\eqref{eq:bin_transport} that reduces the problem to an ODE for the centerline concentration only. Assuming that the source terms do not, to leading order, alter the radial shape of the concentration profile, we set $\mathcal{B}_k + \mathcal{C}_k=0$ and note that every bin concentration must then obey the same equation as the total mean concentration $\tilde{c}$:
 
\begin{equation}
v\,\frac{\partial \tilde{c}}{\partial z} + u\,\frac{\partial \tilde{c}}{\partial r}
= \frac{\mathcal{D}_t}{r}\frac{\partial}{\partial r}\left(r\,\frac{\partial \tilde{c}}{\partial r}\right).
\label{eq:concentration_equation}
\end{equation}

This equation admits a similarity solution provided it is expressed in terms of the total scalar flux injected at the source,
\begin{equation}
Q_i = 2\pi \int_0^{\infty} v(z,r)\, \tilde{c}(z,r)\, r\, dr,
\label{eq:scalar_flux}
\end{equation}
which is independent of $z$. Introducing the nondimensional similarity profile
\begin{equation}
\frac{\tilde{c}_{k}(z,r)}{\tilde{c}_{c,k}(z,0)}
= \frac{\tilde{n}_{k}(z,r)}{\tilde{n}_{c,k}(z,0)} = \phi(\eta),
\label{eq:conc_similarity_ansatz}
\end{equation}
and substituting the mean velocity profile, Eqs.~\eqref{eq:velocity_profile} and \eqref{eq:centerline_laws}, into Eq.~\eqref{eq:concentration_equation}, we obtain an ordinary differential equation governing $\phi(\eta)$. 
\begin{equation}
    \mathcal{D}_t \eta (1+\beta^2 \eta^2)^2 \phi'' + \left ( \mathcal{D}_t (1+\beta^2 \eta^2)^2 + \frac{C_m V_{exit} D \eta^2 (1+\beta^2 \eta^2)}{2} \right) \phi' + C_m V_{exit} D \eta \phi  = 0,
\end{equation}

Writing the eddy viscosity/eddy diffusivity ratio as the turbulent Schmidt number $\text{Sc}_T = \nu_T/\mathcal{D}_t$, and using the constant-eddy-viscosity relation $C_m D V_{exit} = 8\beta^2 \nu_T$ that follows from Eq.~\eqref{eq:centerline_laws}, the equation for $\phi(\eta)$ reduces to a form depending only on $\beta$ and $\text{Sc}_T$. 

\begin{equation}
    Sc_T^{-1} \eta (1+\beta^2 \eta^2)^2 \phi'' + \left ( Sc_T^{-1} (1+\beta^2 \eta^2)^2 + 4 \beta^2 \eta^2 (1+\beta^2 \eta^2) \right) \phi' + 8 \beta^2 \eta \phi  = 0,
\end{equation}

Its solution, monotonically decreasing away from the centerline, is given as
 
\begin{equation}
\phi(\eta) = \frac{1}{\left(1+\beta^2\eta^2\right)^{2\text{Sc}_T}}.
\label{eq:concentration_profile}
\end{equation}
 
Note that this similarity profile is an exact solution of Eq.~\eqref{eq:concentration_equation}
but is applied only approximately to the individual bin concentrations, since the source terms were neglected in its derivation; the approximation enters only through the radial-derivative terms, after which $\eta$ is set to zero. To retain the effect of droplet breakup and coalescence, we return to the full bin transport equation Eq.~\eqref{eq:bin_transport}. Substituting Eqs. \eqref{eq:velocity_profile} and \eqref{eq:conc_similarity_ansatz} into Eq. \eqref{eq:bin_transport} we get

\begin{equation}
V_c(z) f(\eta)
\frac{\partial}{\partial z}
\left[\tilde n_{c,k}(z)\phi(\eta)\right]
=
\left(\mathcal{B}_k + \mathcal{C}_k\right)
+
\frac{1}{r}
\frac{\partial}{\partial r}
\left[
r\mathcal{D}_t
\frac{\partial}{\partial r}
\left[\tilde n_{c,k}(z)\phi(\eta)\right]
\right].
\label{eq:14}
\end{equation}

At the centreline, $\eta=\frac{r}{z-z_v}=0$, where
\begin{equation}
    f(0)=\phi(0)=1,
    \qquad
    \phi'(0)=0,
\end{equation}
the axial derivative becomes
\begin{equation}
\left.
\frac{\partial}{\partial z}
\left(\tilde n_{c,k}\phi\right)
\right|_{\eta=0}
=
\frac{d\tilde n_{c,k}}{dz}.
\end{equation}
For the radial diffusion term,
\begin{equation}
\frac{1}{r}
\frac{\partial}{\partial r}
\left[
r\mathcal{D}_t
\frac{\partial}{\partial r}
\left(\tilde n_{c,k}\phi\right)
\right]
=
\frac{\mathcal{D}_t\tilde n_{c,k}}{(z-z_v)^2}
\left(
\frac{\phi'}{\eta}+\phi''
\right).
\end{equation}
Since
\begin{equation}
    \lim_{\eta\rightarrow0}
    \frac{\phi'}{\eta}
    =
    \phi''(0)
    =
    -4Sc_T\beta^2,
\end{equation}
the radial diffusion term at the centreline becomes
\begin{equation}
\left[
\frac{1}{r}
\frac{\partial}{\partial r}
\left(
r\mathcal{D}_t
\frac{\partial}{\partial r}
(\tilde n_{c,k}\phi)
\right)
\right]_{\eta=0}
=
-\frac{8\beta^2Sc_T\mathcal{D}_t}
{(z-z_v)^2}
\tilde n_{c,k}.
\end{equation}
Using $Sc_T \mathcal{D}_t=\nu_T$ and the constant-eddy-viscosity similarity
relation $8\beta^2\nu_T=C_m D V_{\mathrm{exit}}$, this reduces to
\begin{equation}
\left[
\frac{1}{r}
\frac{\partial}{\partial r}
\left(
r\mathcal{D}_t
\frac{\partial}{\partial r}
(\tilde n_{c,k}\phi)
\right)
\right]_{\eta=0}
=
-\frac{C_m D V_{\mathrm{exit}}}
{(z-z_v)^2}
\tilde n_{c,k}.
\end{equation}
Hence, Eq.~\eqref{eq:14} evaluated at the centreline gives
\begin{equation}
V_c
\frac{d\tilde n_{c,k}}{dz}
=
\left(\mathcal{B}_k + \mathcal{C}_k\right)
-
\frac{C_m D V_{\mathrm{exit}}}
{(z-z_v)^2}
\tilde n_{c,k}.
\end{equation}
The self-similar centreline velocity is
\begin{equation}
\frac{V_{\mathrm{exit}}}{V_c(z)}
=
\frac{1}{C_m}
\left(
\frac{z-z_v}{D}
\right),
\end{equation}
or equivalently,
\begin{equation}
V_c(z)
=
\frac{C_m D V_{\mathrm{exit}}}
{z-z_v}.
\label{eq:Vc_similarity}
\end{equation}
 Dividing the centreline transport equation by $V_c$ and using Eq.~\eqref{eq:Vc_similarity}, we obtain
\begin{equation}
\frac{d\tilde n_{c,k}}{dz}
=
\frac{\left(\mathcal{B}_k + \mathcal{C}_k\right)}{V_c}
-
\frac{\tilde n_{c,k}}{z-z_v}.
\end{equation}
Furthermore,
\begin{equation}
\frac{1}{V_c}
=
\frac{z-z_v}
{C_m D V_{\mathrm{exit}}},
\end{equation}
so that
\begin{equation}
\frac{d\tilde n_{c,k}}{dz}
=
\left(\mathcal{B}_k + \mathcal{C}_k\right)
\frac{z-z_v}
{C_m D V_{\mathrm{exit}}}
-
\frac{\tilde n_{c,k}}
{z-z_v}.
\label{eq:A15_general}
\end{equation}
For $z_v=0$, the above expression  reduces to the 1D-ODE form used in the model

\begin{equation}
 \frac{d\tilde{n}_{c,k}}{dz}
    = \left[\mathcal{B}_k(z,0) + \mathcal{C}_k(z,0)\right]
      \frac{z}{V_{exit}\,C_m\,D}
    - \frac{\tilde{n}_{c,k}(z)}{z}, 
\label{eq:1D_ODE}
\end{equation}
 
where $\mathcal{B}_k$ and $\mathcal{C}_k$ denote, respectively, the net breakup and coalescence source terms evaluated at the centerline. Equation \eqref{eq:1D_ODE} constitutes a system of $M$ coupled ODEs describing the combined effects of turbulent transport,  breakup, and coalescence on the centerline droplet  size distribution. Because breakup and coalescence conserve total droplet volume, $\sum_k (\mathcal{B}_k + \mathcal{C}_k) \Omega_k = 0$. Consequently, the breakup and coalescence source terms do not alter the decay rate of the total centerline concentration $\tilde{c}_c(z) = \sum_k \Omega_k \tilde{n}_{c,k}(z)$ obtained from Eq.~\eqref{eq:centerline_concentration_app}.

Substituting the similarity profiles \eqref{eq:velocity_profile} and \eqref{eq:concentration_profile} into the definition of the scalar flux, Eq.~\eqref{eq:scalar_flux}, and carrying out the radial integration, the resulting expression for the centerline concentration as a function of downstream distance is
 
\begin{equation}
\tilde{c}_{c}(z) = \frac{Q_{i}\,\beta^2 \left(2\text{Sc}_T + 1\right)}{\pi\, V_c(z)\, (z-z_v)^2}.
\label{eq:centerline_concentration_app}
\end{equation}

\section{Breakup Frequency}
\label{appndxB}
Following the kinetic-theory analogy first introduced by \citet{coulaloglou1977description} and later extended by \citet{tsouris1994breakage}, the breakup frequency is modelled as a convolution of the droplet--eddy collision rate and the probability that a given collision imparts sufficient energy to fragment the droplet:
\begin{equation}
\lambda(D_k) = K_s \int_0^{D_k} \frac{\pi}{4}(D_k + \phi_e)^2\, u_e(\phi_e)\,
P_b(D_k, \phi_e)\, dn_e(\phi_e),
\label{eq:breakup_freq_dim}
\end{equation}
where $\phi_e$ denotes the eddy size, $u_e(\phi_e)$ is the characteristic velocity of an eddy of size $\phi_e$, $P_b(D_k,\phi_e)$ is the breakup efficiency defined below, and $K_s$ is a dimensionless empirical coefficient. The upper limit of integration is set to $\phi_e = D_k$ on the physical grounds that eddies larger than the droplet translate rather than deform it. Assuming space-filling eddies, the eddy number density scales as $n_e \propto \phi_e^{-3}$, giving $dn_e = C_1 \phi_e^{-4}\,d\phi_e$ with $C_1$ a constant of order unity.

The velocity scale associated with an eddy of size $r$ is obtained from the second-order longitudinal velocity structure function $\mathcal{F}(r)$, defined as
\begin{equation}
\mathcal{F}(r) = \bigl\langle \bigl[u_L(\mathbf{x} + r\mathbf{e}_L)
- u_L(\mathbf{x})\bigr]^2 \bigr\rangle,
\end{equation}
so that $u_e(r) \sim [\mathcal{F}(r)]^{1/2}$. Classical breakup models \citep{coulaloglou1977description, tsouris1994breakage, zhao2016underwater} invoke Kolmogorov's inertial-range scaling $u_e(r) \sim \varepsilon^{1/3}r^{1/3}$, which becomes inaccurate for droplets approaching the Kolmogorov length scale $\eta_k = (\nu^3/\varepsilon)^{1/4}$. To ensure a physically consistent transition between the viscous and inertial subranges, the Batchelor interpolation formula \citep{batchelor1951pressure} is adopted:
\begin{equation}
\mathcal{F}(r) = C_2\,\varepsilon^{2/3} r^{2/3}
\left[1 + \left(\frac{r}{\gamma_2\,\eta_k}\right)^{-2}\right]^{-2/3},
\label{eq:structure_fn}
\end{equation}
where $C_2 \approx 2.1$ is the Kolmogorov constant and $\gamma_2 = (15C_2)^{3/4} \approx 13$ defines the crossover scale. Equation \eqref{eq:structure_fn} recovers the viscous scaling $\mathcal{F} \propto r^2$ for $r \ll \gamma_2\eta_k$ and the inertial-range scaling $\mathcal{F} \propto r^{2/3}$ for $r \gg \gamma_2\eta_k$.

The breakup efficiency $P_b$ represents the conditional probability that a droplet--eddy collision results in fragmentation. A droplet breaks only if the kinetic energy carried by the colliding eddy exceeds the total energy barrier arising from surface tension and internal viscous stresses \citep{coulaloglou1977description, prince1990bubble}. This is expressed as 
\begin{equation}
P_b(D_k, \phi_e) = \exp\!\left[-\frac{E_\sigma(D_k)
+ E_\nu(D_k)}{E_e(\phi_e)}\right],
\end{equation}
where $E_\sigma$, $E_\nu$, and $E_e$ are the surface-tension resistance energy, the viscous resistance energy, and the eddy kinetic energy, respectively. The surface-tension energy barrier is $E_\sigma(D_k) = 0.0702\pi\sigma D_k^2$, representing the minimum work required to create new interface. The viscous resistance energy, which accounts for the internal dissipation during droplet deformation, is given by \citep{chatzi1989generalized, zhao2016underwater}
\begin{equation}
E_\nu(D_k) = \frac{\pi}{6}\,\alpha\,\varepsilon^{1/3} D_k^{7/3}\,\mu_d
\sqrt{\frac{\rho_c}{\rho_d}},
\end{equation}
where $\alpha \approx 2$ is an empirical constant, $\mu_d$ is the dynamic viscosity of the dispersed phase, and $\rho_c$ and $\rho_d$ are the densities of the continuous and dispersed phases, respectively. The kinetic energy of the colliding eddy, modelled as a sphere of diameter $\phi_e$ with the continuous phase density, is
\begin{equation}
E_e(\phi_e) = \frac{3}{2}\,\frac{\pi}{6}\,\rho_c\,\phi_e^3\,\mathcal{F}(\phi_e).
\end{equation}

Introducing the droplet Reynolds number $Re_k = \varepsilon^{1/3}D_k^{4/3}/\nu$, the droplet Ohnesorge number $Oh_k = \mu_d/\sqrt{\rho_d\sigma D_k}$, and the viscosity--density ratio $\Lambda = (\mu_d/\mu_c)(\rho_c/\rho_d)^{1/2}$, the breakup frequency is cast in the following non-dimensional form:
\begin{equation}
\lambda = \frac{K^*}{\tau_{b,k}} \int_0^1 r_e^{-11/3}(r_e + 1)^2
\left[1 + \left(\frac{r_e\,Re_k^{3/4}}{\gamma_2}\right)^{-2}
\right]^{-1/3}
P_b(Oh_k, Re_k, \Lambda;\,r_e)\,dr_e,
\label{eq:breakup_freq_nd}
\end{equation}
where $r_e = \phi_e/D_k$ is the eddy-to-droplet size ratio, $\tau_{b,k} = \varepsilon^{-1/3}D_k^{2/3}$ is the inertial breakup timescale associated with an eddy of size $D_k$, and $K^*$ is a dimensionless prefactor that absorbs all remaining numerical constants. The non-dimensional breakup efficiency takes the form
\begin{equation}
P_b(Oh_k, Re_k, \Lambda;\,r_e) = \exp\!\left[
-\left(\frac{\Lambda \psi}{Re_k}\right)
\left[1 + \left(\frac{r_e\,Re_k^{3/4}}{\gamma_2}
\right)^{-2}\right]^{\!2/3}
r_e^{-11/3}
\right],
\end{equation}
where $\psi = 0.14\,\Lambda/(Re_k\cdot Oh_k^2) + 0.583$ encapsulates the combined influence of dispersed-phase viscosity and interfacial tension on the breakup resistance. 


\section{Volume-conservative formulation of droplet breakup and coalescence}
\label{appndxC}

This appendix presents the discrete breakup and coalescence source terms used in the population balance model, Eq.~(4.2), and shows that
both conserve the total dispersed-phase volume exactly. The droplet population is divided into $M$ classes with diameters $D_k$ and volumes
$\Omega_k=\pi D_k^3/6$, for $k=1,\dots,M$.

Breakup and coalescence both create droplets whose diameters usually do not match a class diameter: the second (complementary) fragment in breakup, and the merged droplet in coalescence. Both are assigned to the discrete size classes using the volume-conserving rule. Consider a droplet of diameter $D$ with $\Omega_p\le\pi D^3/6\le\Omega_{p+1}$, and let $w_k(D)$ be the fraction of this droplet assigned to class $k$. This fraction is non-zero only for
$k=p$ and $p+1$, and is obtained by linear interpolation in volume:
\begin{equation}
w_{p+1}(D)=\frac{\tfrac{\pi}{6}D^3-\Omega_p}{\Omega_{p+1}-\Omega_p},
\qquad
w_p(D)=1-w_{p+1}(D).
\label{eq:C_weights}
\end{equation}
If $D=D_k$, the entire weight is assigned to class $k$. The weights satisfy the exact identity
\begin{equation}
\sum_k w_k(D)\,\Omega_k=\frac{\pi}{6}D^3 \qquad\text{for any }D,
\label{eq:C_wk_property}
\end{equation}
so that this assignment neither creates nor destroys volume.

\medskip
\noindent Binary breakup of class-$l$ parents occurs at a rate $\lambda(D_l)\,\tilde n_{c,l}$ per unit volume and time, where $\lambda$ is the breakup frequency given in Appendix~B. Each event
produces two daughters with  $\Omega_m+\Omega_c=\Omega_l$. The surface-energy argument of Eqs.~(4.7)-(4.10) gives the probability $\Phi_b(D_m,D_l)$ that the daughter droplet falls in class $m<l$. This probability is normalised so that
\begin{equation}
\sum_{m=1}^{l-1}\Phi_b(D_m,D_l)=1 .
\label{eq:C_norm}
\end{equation}
Once $m$ is chosen, volume conservation fixes the complementary daughter:
\begin{equation}
\Omega_c(m,l)=\Omega_l-\Omega_m,
\qquad
\tilde D(D_m,D_l)=\big(D_l^3-D_m^3\big)^{1/3}.
\end{equation}
Using Eq.~\eqref{eq:C_weights}, the probability that the complementary daughter is assigned to class $k$ is
\begin{equation}
\tilde\Phi_b(D_k,D_l)=\sum_{m=1}^{l-1}\Phi_b(D_m,D_l)\,w_k\!\big(\tilde D(D_m,D_l)\big).
\label{eq:C_tildephi}
\end{equation}
With both daughters included, the breakup source term is
\begin{equation}
\mathcal{B}_k(z,0)=\sum_{l \ge k}\lambda(z,0,D_l)\,\tilde n_{c,l}\,\Phi_{\rm tot}(D_k,D_l)-\lambda(z,0,D_k)\,\tilde n_{c,k},
\qquad
\Phi_{\rm tot}=\Phi_b+\tilde\Phi_b .
\label{eq:C_Bk}
\end{equation}
Here $\Phi_{\rm tot}$ is the full two-daughter distribution of a class-$l$ breakup event, and the last term is the loss of class-$k$ parents. The sum includes $l=k$ because the complementary daughter can be partly assigned back to the parent class. This occurs when the first daughter is small, so that $\tilde D$ lies between $D_{l-1}$ and $D_l$ and, by Eq.~\eqref{eq:C_weights}, a fraction $w_l(\tilde D)$ of it remains in class $l$. Since $\Phi_b(D_l,D_l)=0$, the $l=k$ term contains only this contribution, $\tilde\Phi_b(D_l,D_l)$, and it must be retained for volume to be conserved. From Eqs.~\eqref{eq:C_wk_property} and \eqref{eq:C_norm}, each event returns exactly the parent volume,
\begin{equation}
\sum_k\Phi_{\rm tot}(D_k,D_l)\,\Omega_k=\Omega_l .
\label{eq:C_pereventvol}
\end{equation}
Similarly, since $\sum_k w_k=1$, $\sum_k\Phi_{\rm tot}(D_k,D_l)=2$, i.e.\ each event removes one parent
and adds two daughters. Multiplying Eq.~\eqref{eq:C_Bk} by $\Omega_k$
and summing over $k$, the gain and loss terms cancel for every parent
class, giving
\begin{equation}
\sum_k \mathcal{B}_k(z,0)\,\Omega_k=0 .
\label{eq:C_volcons_B}
\end{equation}
\medskip
\noindent Coalescence of classes $i$ and $j$ produces one droplet of diameter $\tilde D_{ij}=\big(D_i^3+D_j^3\big)^{1/3}$, at the symmetric rate $C(D_i,D_j)\,\tilde n_{c,i}\,\tilde n_{c,j}$ of Eq.~(4.12). This droplet is assigned to discrete neighbouring bins using Eq.~\eqref{eq:C_weights}, and the source term is 
\begin{equation}
\mathcal{C}_k(z,0)=\frac12\sum_{i=1}^{M}\sum_{j=1}^{M} w_k\big(\tilde D_{ij}\big)\,C(D_i,D_j)\,\tilde n_{c,i}\,\tilde n_{c,j}
-\tilde n_{c,k}\sum_{j=1}^{M}C(D_k,D_j)\,\tilde n_{c,j}.
\label{eq:C_Ck}
\end{equation}
In the double sum over all $i,j$, a pair with $i\ne j$ is counted twice, so the factor $\tfrac12$ gives the correct single count. A pair $(i,i)$ is counted once and gives the usual rate $\tfrac12C(D_i,D_i)\tilde n_{c,i}^2$. The last term is the loss of class-$k$ droplets by coalescence with droplets of any class.

To check volume conservation, multiply Eq.~\eqref{eq:C_Ck} by $\Omega_k$ and sum over $k$. Eq.~\eqref{eq:C_wk_property} gives $\sum_kw_k(\tilde D_{ij})\Omega_k=\Omega_i+\Omega_j$. Using the symmetry $C(D_i,D_j)=C(D_j,D_i)$, the gain term becomes
\begin{equation}
\frac12\sum_{i,j}C(D_i,D_j)\,\tilde n_{c,i}\tilde n_{c,j}\,(\Omega_i+\Omega_j)
=\sum_i\Omega_i\tilde n_{c,i}\sum_jC(D_i,D_j)\tilde n_{c,j},
\end{equation}
which is equal to the loss term weighted by $\Omega_i$. Hence
\begin{equation}
\sum_k \mathcal{C}_k(z,0)\,\Omega_k=0
\label{eq:C_volcons_C}
\end{equation}
exactly, for any $\tilde n_{c,i}$. Adding Eqs.~\eqref{eq:C_volcons_B} and \eqref{eq:C_volcons_C} gives $\sum_k[\mathcal{B}_k+\mathcal{C}_k]\Omega_k=0$. Eq.~(4.5) therefore follows directly from the definitions of $\mathcal{B}_k$ and $\mathcal{C}_k$, rather than being a separate condition to be checked
numerically.

\medskip
\noindent In the  numerical implementation, the entries
$\Phi_{\rm tot}(D_k,D_l)$ are collected into a matrix $\boldsymbol\Phi$ (row $k$, column $l$). Since this matrix depends only on the class diameters, it is computed once before the axial marching. Over an axial step $dz$, with $dt=dz/V_c(z)$, the number of breakup 
events of class $l$ per unit volume is 
\begin{equation}
\nu_l=\lambda(D_l)\,\tilde n_{c,l}\,dt.
\label{eq:C_nu}
\end{equation}
The breakup is updated as
\begin{equation}
\Delta\tilde{\mathbf n}_{\rm break}=\big(\boldsymbol\Phi-\mathbf I\big)\,\boldsymbol\nu .
\label{eq:C_matrix_break}
\end{equation}
Since Eq.~\eqref{eq:C_pereventvol} holds for every column, $\boldsymbol\Omega^{T}(\boldsymbol\Phi-\mathbf I)=\mathbf 0$, so this update conserves volume for any $\boldsymbol\nu$.

For coalescence, the number of events between classes $i$ and $j$ in the same step is
\begin{equation}
\nu_{ij}=
\begin{cases}
\tfrac12\,C(D_i,D_i)\,\tilde n_{c,i}^{\,2}\,dt, & i=j,\\[4pt]
C(D_i,D_j)\,\tilde n_{c,i}\,\tilde n_{c,j}\,dt, & i\neq j,
\end{cases}
\label{eq:C_nuij}
\end{equation}
and the update is
\begin{equation}
\Delta\tilde n_{c,k}^{\rm coal}
=\sum_{i\le j} w_k\big(\tilde D_{ij}\big)\,\nu_{ij}
-\tilde n_{c,k}\sum_{j=1}^M C(D_k,D_j)\,\tilde n_{c,j}\,dt.
\label{eq:C_matrix_coal}
\end{equation}
By Eq.~\eqref{eq:C_wk_property}, each $\nu_{ij}$ adds exactly the volume $\Omega_i+\Omega_j$ that is removed from classes $i$ and $j$. This update therefore also conserves volume for any $\nu_{ij}$.

\medskip
\noindent Eqs.~\eqref{eq:C_matrix_break} and \eqref{eq:C_matrix_coal} are applied one after the other at each axial step (Section~4.1). Both conserve the total dispersed-phase volume to machine precision by construction, whatever the time step, the number of classes, or the droplet concentration.


\label{appndxD}
\section{Validation of the model}

\begin{figure}
    \centering
    \includegraphics[scale=0.70]{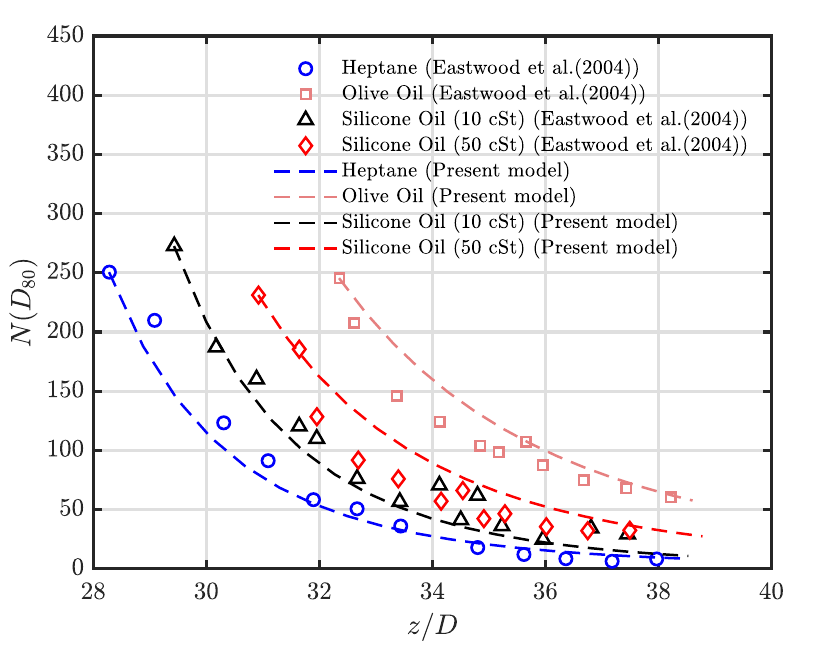}
    \caption{Axial decay of the droplet number density $N(D_{80})$:
    comparison between the 1D ODE model predictions and the experimental data
    of \citet{eastwood2004breakup}.}
    \label{fig:eastwood}
\end{figure}

The applicability of the population balance model has been previously tested against turbulent jet data from \citet{eastwood2004breakup} by \citet{aiyer2019population}. In the present study, we incorporate the coalescence effect and reproduce the corresponding results to further validate the model.  Figure~\ref{fig:eastwood} below compares the axial decay of the number density of droplets at the maximum diameter, $N(D_{80})$, predicted by the population balance model against the measurements of \citet{eastwood2004breakup} for four different dispersed-phase fluids. Here, $D_{80}$ is the maximum diameter such that at least 80\% of the volume of the distribution is contained in droplets smaller than $D_{80}$. The fluid properties and corresponding $D_{80}$ values from Eastwood \emph{et al.} are summarised in Table~\ref{table:oil_properties_eastwood}.
\begin{table}
    \centering
    \caption{Properties of the dispersed phase fluid used by \citet{eastwood2004breakup}}
    \label{table:oil_properties_eastwood}
    \begin{tabular}{c@{\hspace{10pt}}c@{\hspace{10pt}}c@{\hspace{10pt}}c@{\hspace{10pt}}c}
        \toprule
        Fluid              & $D_{80}(\text{mm})$    & $\rho_d~(\text{kg/m$^3$})$      & $\mu_d~(\text{Pa s})$     & $\sigma~(\text{N/m})$\\
        \midrule
        Heptane              & 1.9          & 684                           & 5.0$\times$ $10^{-4}$    & 4.8$\times$ $10^{-2}$ \\
        Olive oil            & 1.91         & 881                           & 7.19$\times$ $10^{-2}$   & 2.0$\times$ $10^{-2}$ \\
        Silicone oil (10 cSt) & 1.92         & 932                           & 9.7$\times$ $10^{-3}$    & 3.5$\times$ $10^{-2}$ \\
        Silicone oil (50 cSt) & 1.87         & 970                           & 5.09$\times$ $10^{-2}$   & 3.7$\times$ $10^{-2}$ \\
        \bottomrule
    \end{tabular}
\end{table}

Following \citet{eastwood2004breakup}, the mean centreline velocity and dissipation rate are modelled as
\begin{equation}
   \frac{V_{exit}}{V_c(z)} = \frac{1}{C_m}\left(\frac{z}{D} - \frac{z_v}{D}
\right),
    \qquad
    r_{1/2}(z) = K\,(z - z_v),
    \label{eq:jet_scaling_appx}
\end{equation}
\begin{equation}
    \varepsilon(z) = C_\varepsilon\,\frac{V_{exit}^3}{D}
    \left(\frac{z}{D} - \frac{z_v}{D}\right)^{-4},
    \label{eq:dissipation}
\end{equation}
with $C_m = 4.08$ and $C_\varepsilon = 36$ (see \citet{eastwood2004breakup}). The virtual origin was found to be $z_v = 5.7D$ by fitting the experimental data with Eq.~\eqref{eq:jet_scaling_appx}, and the spreading parameter was taken as $K = 0.1$.

\citeauthor{eastwood2004breakup} defined $D_{80}$ as a characteristic droplet size whose number density can change only through its own breakup, and cannot increase through the breakup of larger droplets. The corresponding number density $n(D_{80}) = N(D_{80})/V_w$, where $V_w$ is the interrogation window volume, is initially concentrated entirely in the bin corresponding to $D_{80}$, at the location where the oil is injected into the turbulent field.
Its evolution with axial distance is obtained by solving the following equation
\begin{equation}
     \frac{d\tilde{n}_{c,k}}{dz}
    = \left[\mathcal{B}_k(z,0) + \mathcal{C}_k(z,0)\right]
      \frac{(z-z_v)}{V_{exit}\,C_m\,D}
    - \frac{\tilde{n}_{c,k}(z)}{(z-z_v)}, 
    \label{eq:centerline_ode_appx}
\end{equation}
with $D_{80}$ fixed, using fluid properties taken directly from
\citet{eastwood2004breakup}.

For silicone oil of viscosity 10~cSt, the same fluid used as the dispersed phase in the present study, we obtain a breakup coefficient $K^{*} = 0.18$ from this fit.  The corresponding values for heptane, silicone oil (50~cSt), and olive oil are $K^{*} = 0.15$, $0.25$, and $0.25$, respectively. As shown in figure~\ref{fig:eastwood}, the model reproduces the measured axial decay
of $N(D_{80})$ closely for all four fluids, despite their differing viscosity, density, and interfacial tension. Since the present experiments also use 10~cSt silicone oil, we adopt the same value $K^{*} = 0.18$ in the 1D ODE model. This agreement provides direct support for the applicability of the present modelling approach to the flow conditions considered in this study.

\nocite{*}
\bibliographystyle{jfm}
\bibliography{bibl}

\end{document}